\documentclass[twocolumn,twocolappendix]{aastex63}

\usepackage[utf8]{inputenc}
\usepackage[T1]{fontenc}
\usepackage{amsmath,amssymb,bm}
\usepackage{graphicx}
\usepackage{xcolor}
\usepackage{booktabs}
\usepackage{multirow}
\usepackage{makecell}
\usepackage{natbib}
\usepackage{xspace}
\usepackage{lipsum}
\usepackage[switch]{lineno}
\usepackage[toc,page]{appendix}
\usepackage{hyperref}

\def\hst{{\it HST}}
\def\spitzer{{\it Spitzer}}

\newcommand{\smasm}{\rm M_{\odot}}

\newcommand{\smast}{$\rm M_{\odot}$}

\newcommand{\Msun}{\,\mathrm{M}_\odot}     
\newcommand{\Rsun}{\,\mathrm{R}_\odot}     
\newcommand{\Lsun}{\,\mathrm{L}_\odot}     

\newcommand{\Mpc}{\,\mathrm{Mpc}}          

\newcommand{\K}{\,\mathrm{K}}              

\newcommand{\days}{\,\mathrm{days}}

\newcommand{\nifs}{{}^{56}\mathrm{Ni}}
\newcommand{\cofs}{{}^{56}\mathrm{Co}}
\newcommand{\sneii}{SNe~II}
\newcommand{\sneiip}{SNe~II-P}
\newcommand{\sniip}{SN~II-P}
\newcommand{\snii}{SN~II}

\shortauthors{Teixeira et al.}
\shorttitle{SN\,2022acko progenitor constraints}

\begin{document}

\title{SN~2022acko and the Properties of its Red Supergiant Progenitor: Direct Detection, Light Curves, and Nebular Spectroscopy} 


\def\cbpf{\text{a}}

\def\cbpf{Centro Brasileiro de Pesquisas F\'isicas, Rua Dr. Xavier Sigaud 150, 22290-180 Rio de Janeiro, RJ, Brazil}

\def\ciera{Center for Interdisciplinary Exploration and Research in Astrophysics (CIERA) and Department of Physics and Astronomy}

\def\usp{Departamento de Astronomia, Instituto de Astronomia, Geofísica e Ciências Atmosféricas da USP, Cidade Universitária, 05508-090 São Paulo, SP, Brazil}

\author{G. Teixeira}
\address{\cbpf}
\address{\ciera}
\email{gteixeira@cbpf.br}

\author{C. D. Kilpatrick}
\address{\ciera}

\author{C. R. Bom}
\address{\cbpf}

\author{A. Santos}
\address{\cbpf}

\author{P. Darc}
\address{\cbpf}

\author[0000-0002-4449-9152]{K.~Auchettl}
\affil{Department of Astronomy and Astrophysics, University of California, Santa Cruz, CA 95064, USA}
\affil{OzGrav, School of Physics, The University of Melbourne, VIC 3010, Australia}

\author[0000-0002-5045-9675]{Á. Álvarez-Candal}
\affiliation{Instituto de Astrofísica de Andalucía, Glorieta de la Astronomía sn, E18008, Granada, Spain}

\author{R.~J.~Foley}
\affiliation{Department of Astronomy and Astrophysics, University of California, Santa Cruz, CA 95064, USA}

\author[0000-0003-3537-4849]{P. K. Humire}
\affiliation{Departamento de Astronomia, Instituto de Astronomia, Geofísica e Ciências Atmosféricas da USP, Cidade Universitária, 05508-090 São Paulo, SP, Brazil}

\author[0000-0001-6806-0673]{A. L. Piro}
\affiliation{The Observatories of the Carnegie Institution for Science, Pasadena, CA 91101, USA}

\author{C. Rojas-Bravo}
\affiliation{Department of Astronomy and Astrophysics, University of California, Santa Cruz, CA 95064, USA}

\author{C.~Mendes de Oliveira}
\affiliation{Universidade de S\~ao Paulo, IAG, Rua do Matão 1225, São
Paulo, SP, Brazil}
\author{A. Kanaan}
\affiliation{Departamento de F\'isica, \\ Universidade Federal de Santa Catarina, Florian\'opolis, SC, 88040-900, Brazil}

\author{T. Ribeiro}
\affiliation{Rubin Observatory Project Office, 950 N. Cherry Ave., Tucson, AZ 85719, USA}

\author{W. Schoenell}
\affiliation{GMTO Corporation 465 N. Halstead Street, Suite 250 Pasadena, CA 91107, USA}




\begin{abstract}
We present ultraviolet, optical, and infrared observations of the Type~II-P supernova SN~2022acko in NGC~1300, located at a distance of $19.0 \pm 2.9\Mpc$. Our dataset spans $1-350\days$ post-explosion in photometry, complemented by late-time optical spectroscopy covering $200-600\days$, and includes deep pre-explosion imaging. We use this extensive multiwavelength dataset for both direct and indirect constraints on the progenitor system. 
Using the early-time photometry and shock-cooling models, we infer that SN~2022acko likely originated from a red supergiant with a radius of $R \sim 580\Rsun$ and an initial mass of $M \sim 9-10\Msun$. From the radioactive decay tail, we infer a synthesized $\nifs$ mass of $0.014 \pm 0.004\Msun$. We further model nebular-phase spectra using radiative transfer models and nucleosynthesis yields for core-collapse supernovae, which suggest a progenitor initial mass in the range of $10-15\Msun$. Meanwhile, blackbody fitting of the detected pre-explosion counterpart in the F814W and F160W bands indicates a red supergiant with a lower initial mass of approximately $7.5\Msun$. The light curve exhibits a 116~days plateau, indicative of a massive hydrogen-rich envelope, inconsistent with the pre-explosion analysis. We investigated the discrepancy between direct and indirect progenitor mass estimates, focusing on the roles of binary interaction, early-time modeling limitations, and systematic uncertainties in spectral calibration. Our results indicate that the tension among mass estimates likely arises from modeling limitations and flux calibration uncertainties rather than from insufficient data, highlighting the need for more physically realistic models and a deeper understanding of systematic effects.

\end{abstract}



\keywords{
  stars: evolution --- supernovae: general --- supernovae: individual (SN~2022acko)
}



    
    
    

\section{Introduction}
\label{sec:introduction}

Core-collapse supernovae represent the terminal explosions of massive stars with initial masses greater than $8\Msun$ \citep{Woosley1986}. Within this category, Type~II supernovae (SNe~II) are characterized by the presence of prominent hydrogen lines in their spectra, indicating hydrogen-rich progenitor stars \citep{Filippenko1997}. The vast majority of these transient events are believed to result from the explosions of red supergiants (RSGs) \citep{Burrows1995}. Throughout their lifetimes, RSGs undergo successive stages of nuclear burning, synthesizing progressively heavier elements until reaching the iron peak. Once nuclear fusion ceases, outward pressure--sustained by electron degeneracy-- eventually fails due to electron capture and photodissociation, triggering core collapse and a supernova \citep[CCSN;][]{Wheeler_2007}. These explosions play a crucial role in enriching the interstellar medium with heavy elements and regulating star formation in galaxies \citep{Sahijpal_2014, Elmegreen1998}.  However, the detailed physical processes within these stars, especially in the final centuries to years before core collapse and their terminal explosions, remain not fully understood \citep{Davies2020rsgproblem, Fryer2012, Heger2003, Woosley2002}.

Most \sneii\ arise from RSGs with extended hydrogen envelopes, producing a characteristic plateau phase in their light curves that lasts approximately $60-90 \days$, further classifying them as Type II-P supernovae \citep[\sneiip\ ][]{Arcavi2017, Barbon1979, Falk1973}. This plateau phase occurs because, after core collapse, a shock wave propagates through the outer layers of the red supergiant, heating the hydrogen envelope to temperatures of several tens of thousands of Kelvin, fully ionizing the material and making it highly luminous. As the envelope expands and cools, the temperature eventually drops below the hydrogen recombination threshold. At this point, protons and electrons begin to combine into neutral hydrogen. This recombination process releases thermal energy, which slows the decline in luminosity. The recombination front (the layer where hydrogen is recombining) moves inward in mass coordinate (though outward in radius), acting like a photosphere with quasi-constant temperature. Once the entire hydrogen envelope has recombined, the plateau ends. The light curve then drops sharply as it enters the radioactive decay phase, powered by the decay of \(\cofs \rightarrow \nifs\).
In contrast, those lacking a well-defined plateau and instead exhibiting a linearly declining brightness are categorized as Type II-L supernovae \citep{Barbon1979}. Although this classification has been traditionally used, observations suggest that \sneii\ may span a continuum of properties rather than forming two distinct subtypes \citep{Galbany2016, Valenti2016}.

The observed population of \sneii\ with identified RSG progenitors exhibits a zero-age main sequence mass (\( M_\mathrm{ZAMS} \)) distribution spanning approximately 7--19\,M$_\odot$\citep{Smartt2009, Valenti2016, Auchettl_2019, Davies2020rsgproblem}. In contrast, the broader population of RSGs includes stars with initial masses up to $25-30\Msun$ \citep{HumphreysDavidson179, dcb2018, Katsuda2018}. This discrepancy between the initial masses of confirmed \sneii\ progenitors and the full RSG population is known as the \textit{red supergiant problem}. Notably, this discrepancy is also evident in other independent mass indicators, such as nebular spectroscopy of \sneii\ \citep{Fang25}.

Several explanations have been proposed to account for the RSG problem, one possible explanation is that some massive RSGs may undergo core collapse and collapse directly into a black hole without producing an observable supernova, resulting in so-called ``failed supernovae'' \citep{Sukhbold_2016}.  Another explanation is that a small fraction of RSGs may exhibit lower luminosities in their final stages due to extreme variability \citep{Soraisam2018, Levesque2020, Jencson2022, Beasor_2025}. 

Alternatively, the RSG problem may arises from observational biases, particularly the limited number of supernovae with confidently identified progenitor stars \citep{Davies2020rsgproblem}. In such cases, progenitor mass estimates derived from pre-explosion images may be systematically biased by factors such as variability or the absence of infrared observations, which can obscure or underestimate the intrinsic luminosity. As a result, independent methods for estimating progenitor masses (such as nebular spectroscopy, light curve modeling, or environment analysis) are critical for analyzing whether our detections yield accurate and consistent mass constraints.  Resolving the RSG problem requires a larger sample of  supernovae observations with well-characterized progenitor systems. Such efforts are essential not only for clarifying the fate of massive stars but also for advancing our broader understanding of the \sneiip\ population.

The ideal scenario for connecting SN explosions to their respective progenitor stars and estimating their mass involves having archival data that directly captures the progenitor in pre-explosion imaging \citep[e.g., ][]{VanDyk2017}. Such data enable mass estimates through template fitting by comparing the observed luminosity with stellar evolution models across a range of initial stellar masses. The best-fitting template corresponds to a set of stars whose predicted luminosities are consistent with the observed value, thereby constraining the progenitor’s initial mass. In the case of \sneiip\, which are known to originate from RSGs, the progenitor mass can also be estimated by fitting RSG spectral energy distribution (SED) models to the observed SED, particularly when multi-band photometry is available. While these methods offer direct constraints on progenitor mass, they are often not feasible--particularly for more distant SNe ($\gtrsim 28\Mpc$), as obtaining photometry of resolved massive stars becomes increasingly difficult beyond this range \citep{Smartt_2015}.

In addition to using direct detections, several indirect methods can be used to infer progenitor properties from the observed \sneiip. For example, on the absence of optically thick circumstellar material (CSM) around the progenitor, the early-time emission is primarily shaped by the physical structure of the progenitor star, particularly its radius and envelope mass. When the shock wave generated by the core collapse reaches the stellar surface, in the so-called shock breakout phase, it releases a brief burst of high-energy radiation. This is immediately followed by a cooling phase, during which the hot, expanded envelope radiates thermal energy as it cools and becomes progressively more transparent \citep{Morag2023, Waxman_2017, Sapir_2017}. The duration, shape, and brightness of this early light curve are directly tied to the progenitor’s properties.

For \sneiip, the plateau phase of the light curve is powered by hydrogen recombination in the extended hydrogen-rich envelope \citep{Kasen_2009, Popov1993}. When recombination ceases and the ejecta become optically thin, the explosion transitions into the nebular phase. During this stage, the spectrum is dominated by emission lines from elements synthesized during the progenitor’s lifetime and the explosion itself \citep{ Jerkstrand2017, dessart2021}. In this phase, prominent lines such as [O\,\textsc{i}] at 5577, 6300, and 6364\,\AA\ and [Ca\,\textsc{ii}] at 7292 and 7324\,\AA\ correlate with the total ejecta mass of these elements \citep{JerkstrandSN2004et}. By comparing observed nebular-phase spectra with theoretical non-local thermodynamic equilibrium radiative transfer (NLTERT) models informed by nucleosynthesis yields, one can estimate the progenitor’s initial mass by matching the synthetic spectra to the observed features \citep[e.g.,][]{jerkstrend_12m,  Tinyanont2021, kilpatrick2023type}. 

Beyond the initial mass, other SN observables can also be linked to progenitor properties. For instance, studies by \citet{Childress2015} and \citet{Valenti2016} suggest that the duration and luminosity of the plateau phase, as well as the amount of synthesized $\nifs$ correlate with the progenitor’s mass. These correlations are often interpreted through empirical mappings derived from \sneii\ with direct progenitor detections, such as those presented in \citet{eldridge2019supernova}. These photometric indicators thus provide additional constraints on the nature of the progenitor system, complementing our spectroscopic analyses.


In this paper, we analyze SN~2022acko, discovered in the galaxy NGC~1300 by the Distance Less Than 40 Mpc \citep[DLT40, ][]{Tartaglia2018} Survey on 2022 December 6. SN~2022acko was classified as a \sniip\ within the first 24\,hr of detection, based on spectral data obtained with the Yunnan Faint Object Spectrograph and Camera (YFOSC) \citep{Wang_2019}, as reported by \citet{li2022}.

Here, we present new optical and ultraviolet photometry of SN~2022acko spanning 1--350 days after detection, complemented by deep pre-explosion imaging previously reported by \citet{vandyk_2023}, as well as additional late-time spectroscopy covering approximately $200-600\days$ post-explosion. This study focuses on constraining the progenitor’s initial mass using both direct estimates from pre-supernova imaging and indirect inferences based on the supernova’s photometric and spectroscopic evolution. Throughout this paper, we adopt a redshift of $z = 0.00526$ \citep{Springob2005} and a distance of $19.0 \pm 2.9$\,Mpc \citep{Anand2020} for NGC\,1300.

This paper is structured as follows. In Section~\ref{sec:observations}, we describe the acquisition of the photometric and spectroscopic observations of SN~2022acko presented in this work. Section~\ref{sec:progenitor} details the detection and characterization of the pre-explosion counterpart of SN~2022acko. Section~\ref{sec:analysis} presents our estimates of the progenitor system based on SN observables, including shock-cooling modeling, pseudo-bolometric light curve construction, and nebular-phase spectral fitting. In Section~\ref{sec:discussion}, we discuss the implications of our estimates for the progenitor properties and evaluate the role of mass loss during the progenitor’s lifetime. Finally, Section~\ref{sec:conclusions} summarizes our main findings.

\section{Observations}
\label{sec:observations}


    \subsection{Pre-explosion data}\label{sec:observations-pre}

    We analyzed all available {\it Hubble Space Telescope} ({\it HST}) and {\it Spitzer Space Telescope} ({\it Spitzer}) imaging obtained at the site of SN~2022acko in NGC~1300.  The \hst\ imaging was obtained from 6 January 2001 to 4 January 2020, or $\approx$22 to 3~yr prior to explosion, using the Wide Field Planetary Camera 2 (WFPC2), Advanced Camera for Surveys (ACS), and Wide Field Camera 3 (WFC3) as summarized in Table~\ref{tab:candidate_photometry}.  We downloaded all calibrated \hst\ image frames from the Mikulski Archive for Space Telescopes\footnote{\url{https://mast.stsci.edu/}} using our automated pipeline {\tt hst123} \citep{Kilpatrick2021}.  Each image was optimally aligned using {\tt TweakReg} and individual epochs and filters were separately drizzled using {\tt astrodrizzle}.  We then performed photometry across the calibrated and aligned frames using {\tt dolphot} \citep{dolphot} with the F814W image as a reference frame.  We saved all photometry of point-like sources for analysis, which is further described below in Section~\ref{sec:progenitor}.

    We similarly processed the {\it Spitzer} imaging from the Infrared Array Camera (IRAC) using our custom pipeline described in \citep{kilpatrick2023type,forwardmodel}.  There were four separate epochs of IRAC imaging obtained from 2009--2011, which we combined into two stacks for each filter using {\tt mopex} \citep{mopex}.  We performed photometry of point sources in each frame using the instrumental point-spread function (PSF) from \spitzer.

    We detect a single point-like counterpart close to the reported site of SN~2022acko in \hst/F814W and F160W imaging, and we report the brightness of this source and upper limits in all remaining bands in Table~\ref{tab:candidate_photometry}.  We assess whether this candidate is likely associated with SN~2022acko in Section~\ref{sec:progenitor}.

    \begin{table}
    \centering
    \begin{tabular}{lccc}
    \hline
    MJD        & Instrument & Filter  & Brightness (AB mag) \\ \hline
    51915.55779 &  WFPC2     &   F606W    &    $>$25.99 \\
    53269.61735 &  ACS/WFC   &   F555W &    $>$26.86 \\
    53269.62340 &  ACS/WFC   &   F435W &    $>$27.43 \\
    53274.49244 &  ACS/WFC   &   F814W &    26.24$\pm$ 0.07      \\
    53274.49845 &  ACS/WFC   &   F658N &    $>$24.9      \\
    58053.50298 &  WFC3/IR   &   F160W &    24.42$\pm$0.03      \\
    58852.38389 & WFC3/UVIS  &   F336W &    $>$26.33      \\
    58852.44509 &  WFC3/UVIS &   F275W &     $>$25.74      \\
    \hline\hline 
    --          & {\it Spitzer}/IRAC & Ch 1 & $>$22.00 \\
    --          & {\it Spitzer}/IRAC & Ch 2 & $>$22.05 \\ \hline
    \end{tabular}
    \caption{ Photometry of the pre-explosion counterpart to SN~2022acko from \hst\ and \spitzer\ imaging.\label{tab:candidate_photometry}}
    \end{table}
    
    \subsection{High-resolution imaging}

    We obtained a high-resolution image of SN~2022acko with the Gemini South Adaptive Optics Imager \citep[GSAOI;][]{GSAOI} in $H$-band on 11 January 2023, roughly 38~days post-explosion.  All imaging was obtained in conjunction with the Gemini Multi-conjugate Adaptive Optics System \citep[GeMS;][]{GeMS} in laser guide star mode, with 43$\times$30~s images and resulting in an average full-width at half-maximum (FWHM) of the SN~2022acko point spread function (PSF) of 100~mas.  We processed all GSAOI imaging using Gemini {\tt pyraf} methods \citep{gemini_pyraf}, including pixel-level calibration using dark and flat-field frames, sky subtraction from sky frames obtained 3\arcmin\ from SN~2022acko and obtained in conjunction with the science frames, distortion correction with {\tt disco\_stu}\footnote{\url{https://www.gemini.edu/sciops/data/software/disco_stu.pdf}}, and image stacking to a common, undistorted image frame using {\tt SWarp} \citep{swarp}.  SN~2022acko and several dozen unresolved astrometric calibrators in NGC~1300 are clearly detected in the final image stack, which we discuss further in Section~\ref{sec:progenitor}.

    \subsection{Optical/UV observations}
    We acquired optical photometry data from \citet{Bostroem_2023}, covering the first 150 days after explosion, obtained using several instruments. This dataset includes \textit{UBgVri}-band observations from the Las Cumbres Observatory robotic telescope network \citep{Brown_2013}, \textit{co}-band data from the Asteroid Terrestrial-impact Last Alert System (ATLAS; \citealt{Tonry_2018}), and \textit{BgVri} and \textit{Open}-filter observations from the 0.4 m PROMPT telescopes as part of the DLT40 program \citep{Tartaglia2018}. 
    
    Additional ultraviolet and optical observations were obtained with the Ultraviolet/Optical Telescope (UVOT) aboard the Neil Gehrels Swift Observatory, in the \textit{UVW2}, \textit{UVM2}, \textit{UVW1}, \textit{U$_S$}, \textit{B$_S$}, and \textit{V$_S$} bands \citep{swift_telescope}.
    
    Furthermore, we supplemented the light curve dataset with \textit{griz}-band imaging obtained using the 0.8m T80-South robotic telescope located at the Cerro Tololo Inter-American Observatory (CTIO), Chile, as part of the S-PLUS Transient Extension Program \citep[STEP; ][]{santos2024s}. STEP emerges as a time-domain branch of the on-going Southern Photometric Local Universe survey \citep{MendesDeOliveira+19}, covering 8.300 deg$^{2}$ in the southern hemisphere. Early observations were acquired starting 13 days after the discovery date on December 19, 25 and 30 of 2022, while late-time imaging was obtained between 300 days after explosions. We performed forced photometry at the supernova location using a modified version of {\tt DoPhot}, measuring AB magnitudes after image subtraction using PAN-STARRS DR2 images as templates.
    
\begin{figure*}
        \centering
        \includegraphics[width=\linewidth]{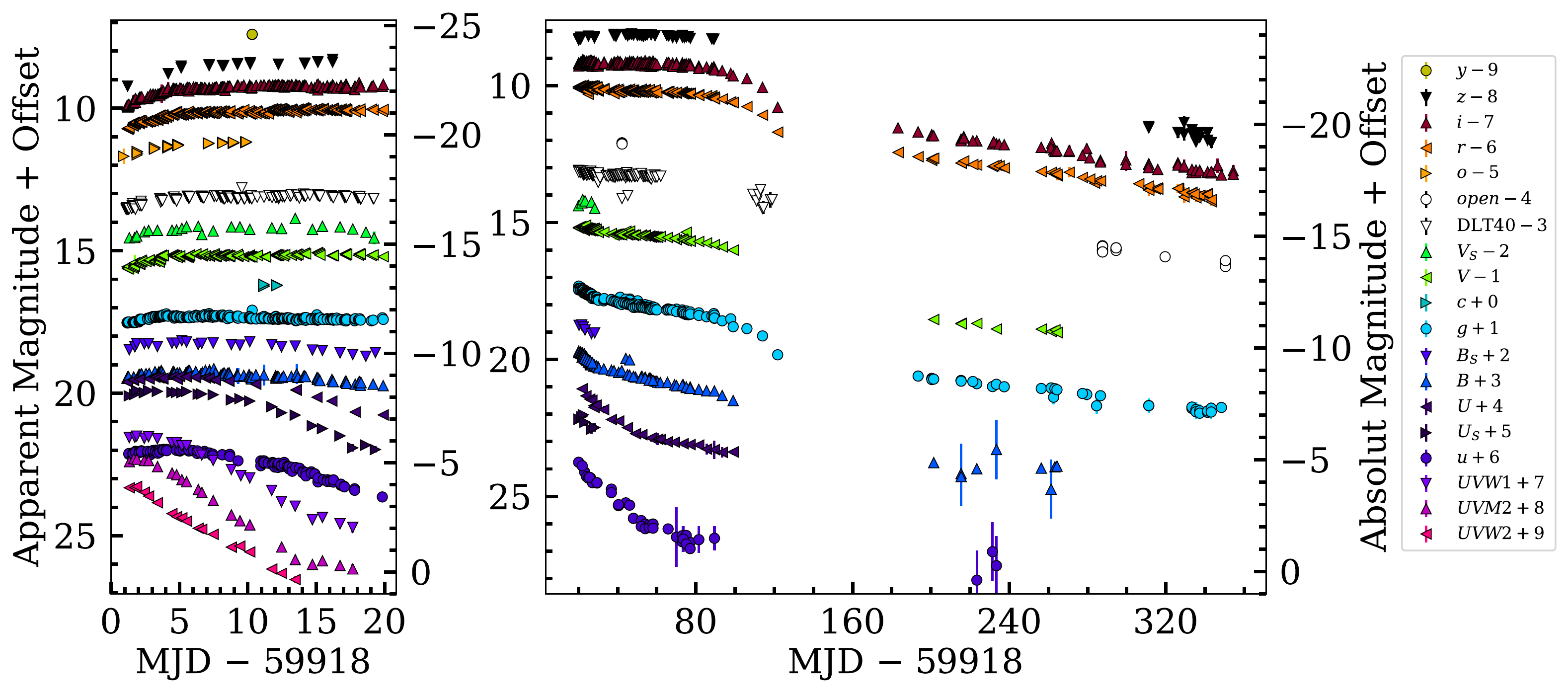}
        \caption{{Apparent and absolute magnitudes of SN~2022acko over the first 350 days of observation. The left panel highlights the early-time evolution (first 20 days), while the right panel displays the later phases. Apparent magnitudes are shown on the left y-axis, and the corresponding absolute magnitudes are overlaid on the right y-axis. The y-axes labels and the legend include offset values for a better visual representation of the data.}} 
        \label{fig:data_discplacement}
    \end{figure*}

    An additional set of late-time photometric data was also obtained 200 days after explosion using the same filter configuration with the Las Cumbres Observatory network, and in $G$ and \textit{Open} filters through the DLT40 survey. These datasets provide valuable constraints on the fading light curve, enhancing our temporal coverage during the decline phase. The novel light-curve information not included in \cite{Bostroem_2023} can be found in Appendix~\ref{sec:appendix_photometry} . All photometric observations of SN~2022acko used in this work are shown in Figure~\ref{fig:data_discplacement}.

    \subsection{Spectroscopy}
    \label{sec:observation-spec}
    We observed SN~2022acko with the Keck-I and Keck-II 10\,m telescopes at Mauna Kea Observatory, Hawaii.  We used the DEIMOS spectrograph \citep{deimos} on Keck-II on 2023 Sep 5 (275~days post-explosion) and the LRIS spectrograph \citep{lris} on Keck-I 2024 Jan 6 and 2024 Aug 9 (396 and 612~days post-explosion, respectively).  DEIMOS was used in a low-resolution 600ZD grating mode with a central wavelength approximately at 7500~\AA\ and flux calibrated with the standard star BD~28+4211.  Similarly, we used the B400/3400 grism and R400/8500 grating in conjunction with the d560 dichroic on LRIS, covering approximately 3000--10500~\AA.  We flux calibrated all epochs using spectra of BD~28+4211 obtained on the same night as the SN~2022acko observations.  All data were processed with {\tt pypeit} v1.16 \citep{pypeit}, using dome flats and arc lamp exposures obtained on the same night and in the same instrumental setup as each science frame.  We performed optimal coaddition of the one-dimensional spectra and telluric correction using {\tt pypeit} atmospheric grids for Maunakea.  The final calibrated spectra are analyzed in Section~\ref{sec:analysis}.
    


\section{The progenitor star of SN 2022\MakeLowercase{acko}}
\label{sec:progenitor}

\subsection{A pre-explosion counterpart to SN~2022acko}

\begin{figure}
    \centering
\setlength{\fboxsep}{-0.45pt}
\setlength{\fboxrule}{1.0pt}
\includegraphics[width=0.48\textwidth]{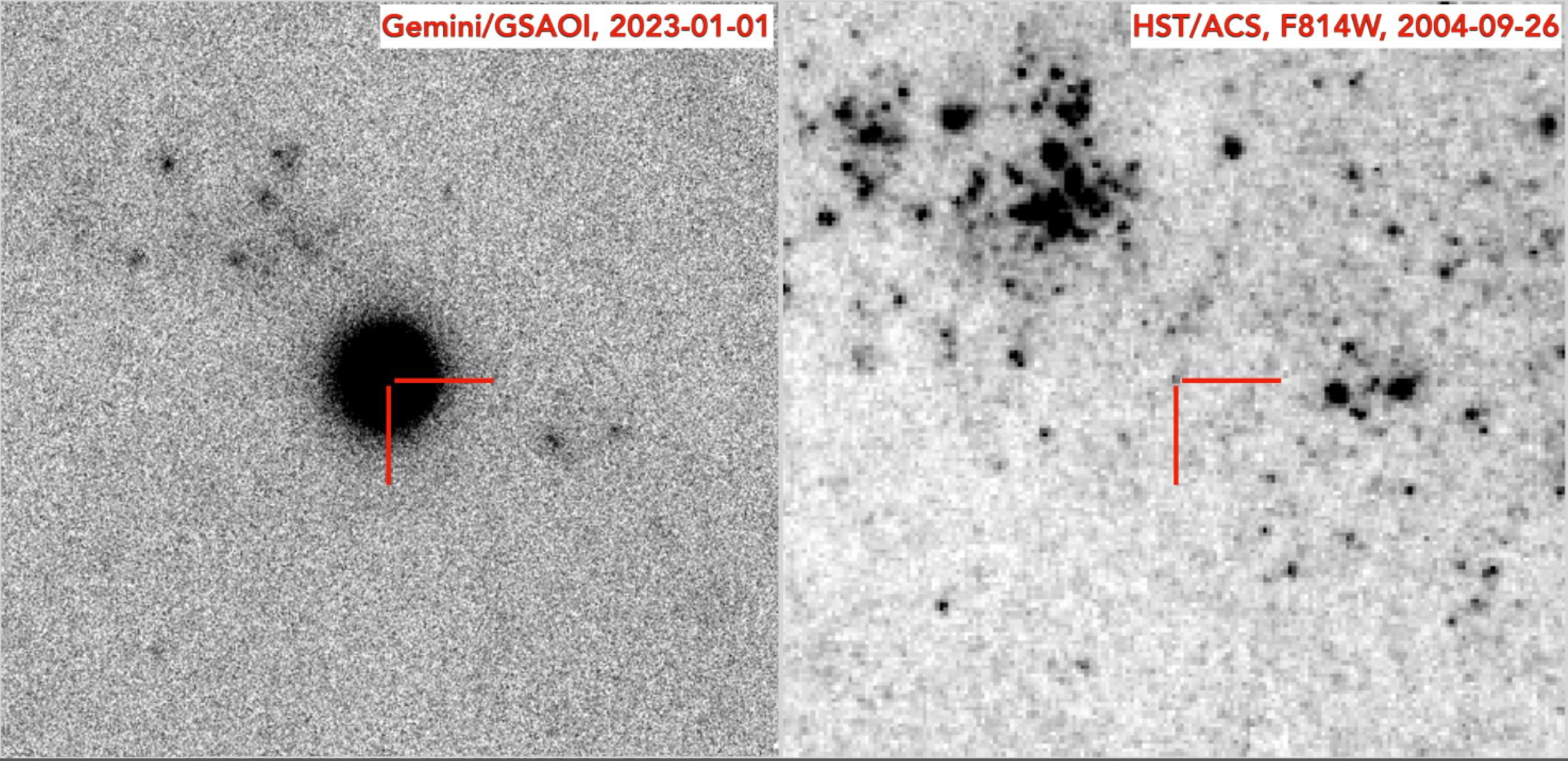}
    \caption{The site of SN~2022acko from 1 January 2023 as observed in high-resolution GSAOI $H$-band imaging ({\it left}) compared with the same site in pre-explosion ACS F814W imaging from 26 September 2004 ({\it right}).  We identify a single, unblended, point-like source in the pre-explosion emission (shown at the position of the red crosshairs) that is astrometrically consistent with the supernova.} 
    \label{fig:progenitor_comp}
\end{figure}

We aligned our GSAOI $H$-band image of SN~2022acko to pre-explosion \hst\ imaging in F814W in order to determine whether a counterpart was present in pre-explosion imaging at the site of the supernova.  We identified 27 sources present in both the GSAOI image and the \hst\ imaging that appear point-like in both frames (Figure~\ref{fig:progenitor_comp}).  Following methods in \citet{kilpatrick2023type}, we use these sources to align the two images frames in the following way.   We first split the sample of common alignment sources in half and calculate an alignment solution that aligns the Gemini/GSAOI$\rightarrow$\hst\ image frame.  We then estimate the systematic uncertainty from our solution by comparing the predicted coordinates of the remaining half to the sample using this solution.  Repeating this procedure 1000 times, we take the average dispersion in right ascension and declination between the calculated GSAOI source coordinates and known \hst\ source coordinates as the systematic uncertainty in our solution, which is approximately $\delta{\rm R.A.}=0.02\arcsec$ and $\delta{\rm Decl.}=0.03\arcsec$.  We take the best-fitting alignment solution as the one calculated from all 27 sources.

Next, we determine where SN~2022acko is located in pre-explosion imaging based on the alignment solution estimated above.  Due to the extremely high signal-to-noise detection of SN~2022acko in our GSAOI imaging, the uncertainties in our centroid from that detection are negligible compared with our alignment uncertainties (e.g., $<$1~mas in the GSAOI frame).  Thus, we find that the resulting position of SN~2022acko is coincident with a single, unblended point-source detected at $\approx$14$\sigma$ significance in the F814W image to within $\approx$0.01\arcsec (nominally 0.4$\sigma$ of the astrometric uncertainty), well within the systematic uncertainties calculated above.  There are no other point-like sources detected at the $>$3$\sigma$ level within $\approx$0.2\arcsec\ (8$\sigma$ of the astrometric uncertainty) of SN~2022acko. This pre-explosion counterpart identification is consistent with the independent alignment performed using {\it JWST}/NIRCam imaging reported in \citet{vandyk_2023}.

We estimate the probability of chance coincidence between SN~2022acko and the pre-explosion counterpart by considering that we detect 254 point-like sources at $>$3$\sigma$ significance in the F814W frame within 10\arcsec\ of SN~2022acko.  This means that there is a $\approx$1.3\% chance that SN~2022acko would have intersected with a random point source to within 3$\sigma$ astrometric significance by chance, which we take as a conservative estimate of the probability of chance coincidence.  While nominally a low significance, this residual low likelihood of a chance detection reinforces the need for future \hst\ follow up to determine whether the pre-explosion counterpart has in fact disappeared, confirming its association with SN~2022acko.

Finally, with the detection of a single, high-significance counterpart to SN~2022acko in our F814W image, we align this frame to the remaining \hst\ and \spitzer\ imaging.  The alignment uncertainties are generally extremely small ($\approx$a few mas) in \hst\ imaging, but they approach 0.01\arcsec\ in both \spitzer/IRAC Channels 1 and 2.  We obtain a detection of the counterpart in SN~2022acko as previously reported in \citet{vandyk_2023}, but no detections in any other \hst\ or the \spitzer\ bands. As this source is aligned with the position of the counterpart constrained from the discovery coordinate and reported in Table~\ref{tab:candidate_photometry}, we use those data in the analysis described below.

\subsection{Characterization of the Pre-explosion Counterpart}\label{sec:progenitor_sed}

\begin{figure}
    \includegraphics[width=0.49\textwidth]{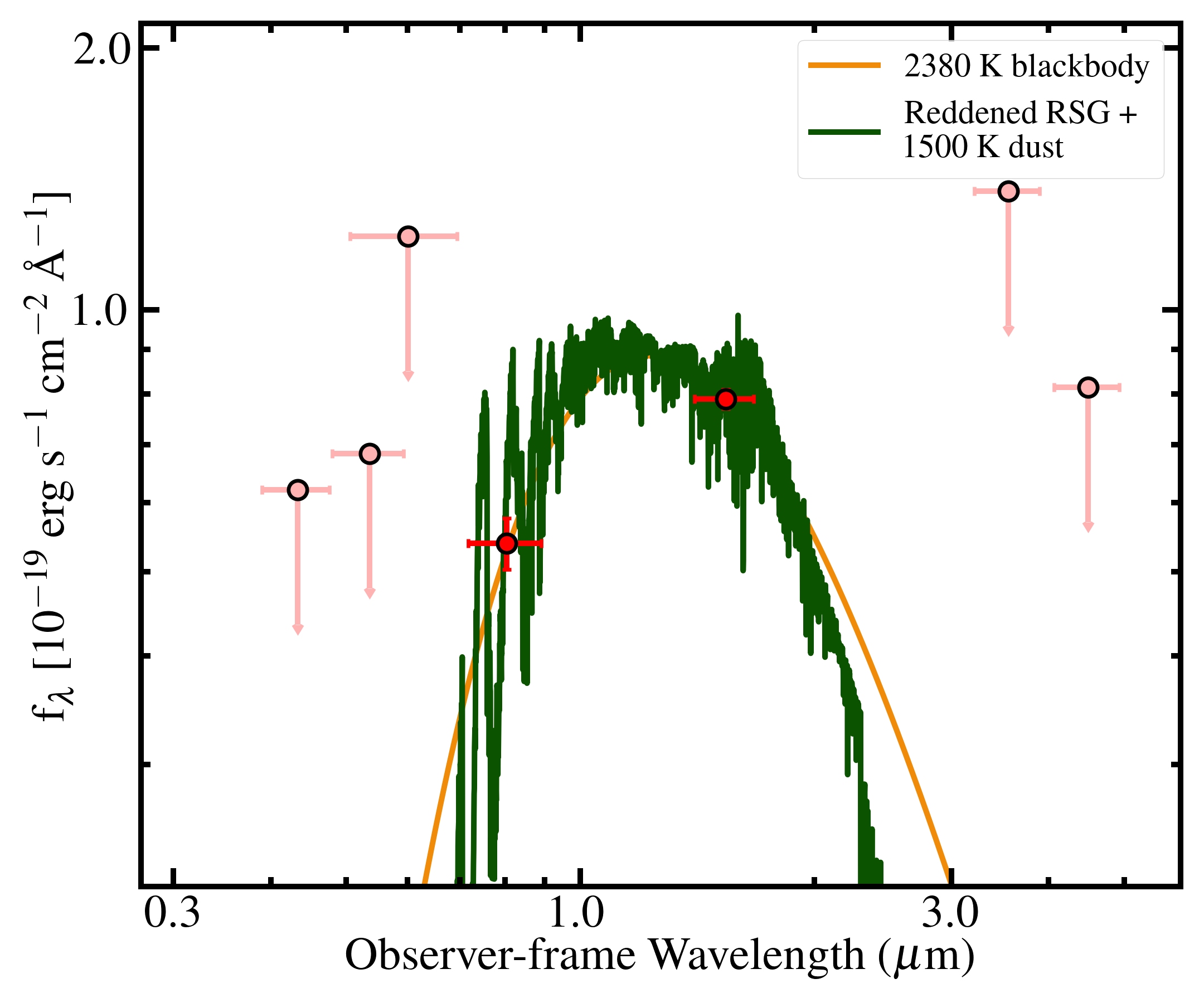}[h]
    \caption{Photometry of the SN~2022acko pre-explosion counterpart in \hst\ F814W and F160W (red circles) as described in Section~\ref{sec:observations-pre}.  Upper limits from other \hst\ and \spitzer\ bands are shown in pink.  We fit these data using two models described in Section~\ref{sec:progenitor_sed}; a pure blackbody with a temperature of 2380~K and luminosity of $\log(L/\Lsun)=4.27$ and a more realistic MARCS RSG with added circumstellar reddening from a shell of $1500$~K dust and a total luminosity of $\log(L/\Lsun)=4.3$.}\label{fig:sed}
\end{figure}

With the detection of the SN~2022acko pre-explosion counterpart in F814W and F160W imaging as well as limits from F336W to IRAC Channel 2, we characterize the potential luminosity and spectral type of the putative counterpart using methods described in \citet{kilpatrick2023type} and \citet{Kilpatrick_2023}.  We compare these data using a MCMC fitter using both a simple blackbody model (i.e., fit with $T_{\rm eff}$ and $\log(L/\Lsun)$) and a grid of MARCS \citep{Gustafsson2008} stellar SEDs combined with an opacity simulating a RSG stellar wind described in the papers above.  The model is described using the intrinsic RSG luminosity ($L$), the effective temperature of the RSG photosphere ($T_{\rm eff}$), a line-of-sight opacity through the RSG wind in $V$-band ($\tau_{V}$), and a temperature describing cool blackbody emission from stellar light reprocessed in any dust present in the wind ($T_{\rm dust}$).  We incorporate line-of-sight extinction from the Milky Way, but due to the lack of any apparent host extinction in the light curves and spectra of SN~2022acko, we do not include additional extinction from NGC\,1300.  The results of our best-fitting SED are presented in Figure~\ref{fig:sed}.  For comparison, we also show the results of the best-fitting model assuming pure blackbody emission between F814W and F160W.

The best-fitting blackbody results in a photosphere with effective temperature 2380~K and $\log(L/\Lsun)=4.27\pm0.18$, which is significantly cooler than realistic RSG SEDs from the MARCS models.  We note that as these data are fit to two photometric points, there could be significant systematic uncertainties on these quantities due to overfitting.  Assuming that the underlying source is dominated by a single RSG photosphere and given the lack of significant line-of-sight extinction, we infer that there is some additional source of reddening in the local environment around the SN~2022acko progenitor system that our RSG model may capture.  

Moreover, the implied luminosity is consistent with a 7.5$\pm$0.5~$M_{\odot}$ RSG following models in \citet{mist}, which is extremely low for a bare RSG photosphere.  This suggests that either the SN~2022acko progenitor system is anomalously low mass for a \sniip\ progenitor star or we are missing a significant fraction of the emission from this source at wavelengths beyond F814W and F160W.  We note that this conclusion is consistent with findings in \citet{Van_Dyk_2023}, who compare to BPASS \citep{BPASS} models with luminosities as low as $\log(L/\Lsun)=4.3$ and initial mass $7.7~\Msun$.

We instead turn to our MARCS RSG models to determine the most luminous the SN~2022acko progenitor system could be once we incorporate the {\it Spitzer}/IRAC Channels 1 and 2 limits.  Combining systematic uncertainty in the model and uncertainty in the distance, the maximum luminosity model that is consistent with our photometry at 1$\sigma$ is $\log(L/\Lsun)\approx4.5$.  

\section{Physical properties of SN 2022\MakeLowercase{acko}}\label{sec:section3_physicalproperties}\label{sec:analysis}
\subsection{Early phases and shock cooling modeling}\label{subsec:sc}
    The early light curves of SNe are dominated by shock cooling (SC) emission, which constrains the progenitor structure and energy deposited in the outer progenitor envelope during shock breakout. We analyze and compare two versions of SC modeling developed by \citet{Sapir_2017} (SW17) and \citet{Morag2023} (MSW23). The MSW23 is an modified version of SW17, accounting for line blanketing in the UV bands at the earlier phases by using a modified spectral energy distribution (SED). 

    Both models can be described by the following set of parameters: the stellar radius $R$, its shock velocity profile $v_{\rm sh}$, the ejecta mass $M$, the hydrogen envelop mass $M_{\rm env}$ and the time of explosion $t_0$. In the SC formalism, the product $f_{\rho}M$ consistently appears as a single term, where  $f_{\rho}$ is a numerical factor that depends on the assumed polytropic density profile of the envelope. In that sense, we treat $f_{\rho}M$ as a single effective parameter in our analysis.  

    The models were implemented using the \texttt{lightcurve\_fitting} package \citep{griffin_hosseinzadeh_2023_8049154}, and the parameters were obtained by executing MCMC routines. This methodology has also been applied to other SNe, including SN 2021yja, SN 2023ixf, and SN 2020jfo \citep{Hosseinzadeh2022_2021yja, Hosseinzadeh_2023, kilpatrick2023type}. To enforce a physically meaningful constraint — that the ejecta mass must be at least on the order of the envelope mass — we modified the log-likelihood function to discard samples where $f_{\rho_M} / M_{\rm env} < 0.1$. This ensures that only models where the parameter $f_{\rho}$ corresponds to physically plausible ejecta configurations are retained in the posterior distribution, in line with expectations from radiation-hydrodynamic simulations for \citep[see Figure~5 in ][]{Sapir_2017}.
    
    We ran the MCMC with 32 walkers for 5000 burn-in steps and additional 5000 steps for sampling. We present the best-fit values from both modelings and their respective prior distributions in Table~\ref{tab:shock_fit}. In this run, we included the intrinsic scatter parameter $\sigma$, which increases the observed error bars by a factor of $\sqrt{1+\sigma}$.

    \begin{table*}
    \centering
    \begin{tabular}{c c c c c c c c } 
                & & \multicolumn{3}{c}{Prior} & \multicolumn{2}{c}{Best fit values} & \\
            \cmidrule(r){3-5} 
            \cmidrule(r){6-7} 
            Parameter & Variable & Shape & Min. & Max & MSW23 & SW17 & Units \\
             \hline
            
            description & $v_\mathrm{s*}$  & Uniform & 0 & 10.0 &
            $1.12^{+0.06}_{-0.06}$ & $0.7^{+0.1}_{-0.1}$ & $10^3$~km~s$^{-1}$ \\
            
            description & $M_\mathrm{env}$  & Uniform & 0 & 10.0 &
            $2.0^{+0.4}_{-0.3}$ & $1.3^{+0.2}_{-0.2}$ & $ \Msun$ \\
            
            description & $f_\rho$ M  & Uniform & $0.1\times M_\mathrm{env}$ & 100 &
            $0.21^{+0.04}_{-0.03}$ & $6.0^{+9.0}_{-5.0}$ & $ \Msun$ \\
            
            description & $R$  & Uniform & 0 & 14374 &
            $580.0^{+20.0}_{-20.0}$ & $1000.0^{+100.0}_{-200.0}$ & $\Rsun$ \\
            
            description & $t_0$  & Uniform & 59910.94 & 59920.44 &
            $59917.44^{+0.06}_{-0.06}$ & $59916.2^{+0.1}_{-0.1}$ & MJD \\
            
            description & $\sigma$  & LogUniform & 0 & $10^{2}$ &
            $6.8^{+0.1}_{-0.1}$ & $6.3^{+0.1}_{-0.1}$ & Dimensionless \\
            
            \hline
            \end{tabular}
            \caption{Parameters priors and best fit values encountered by the shock4 model.
            }
        \label{tab:shock_fit}
    \end{table*}
           
\begin{figure}
        \centering
        \includegraphics[width=.4\textwidth]{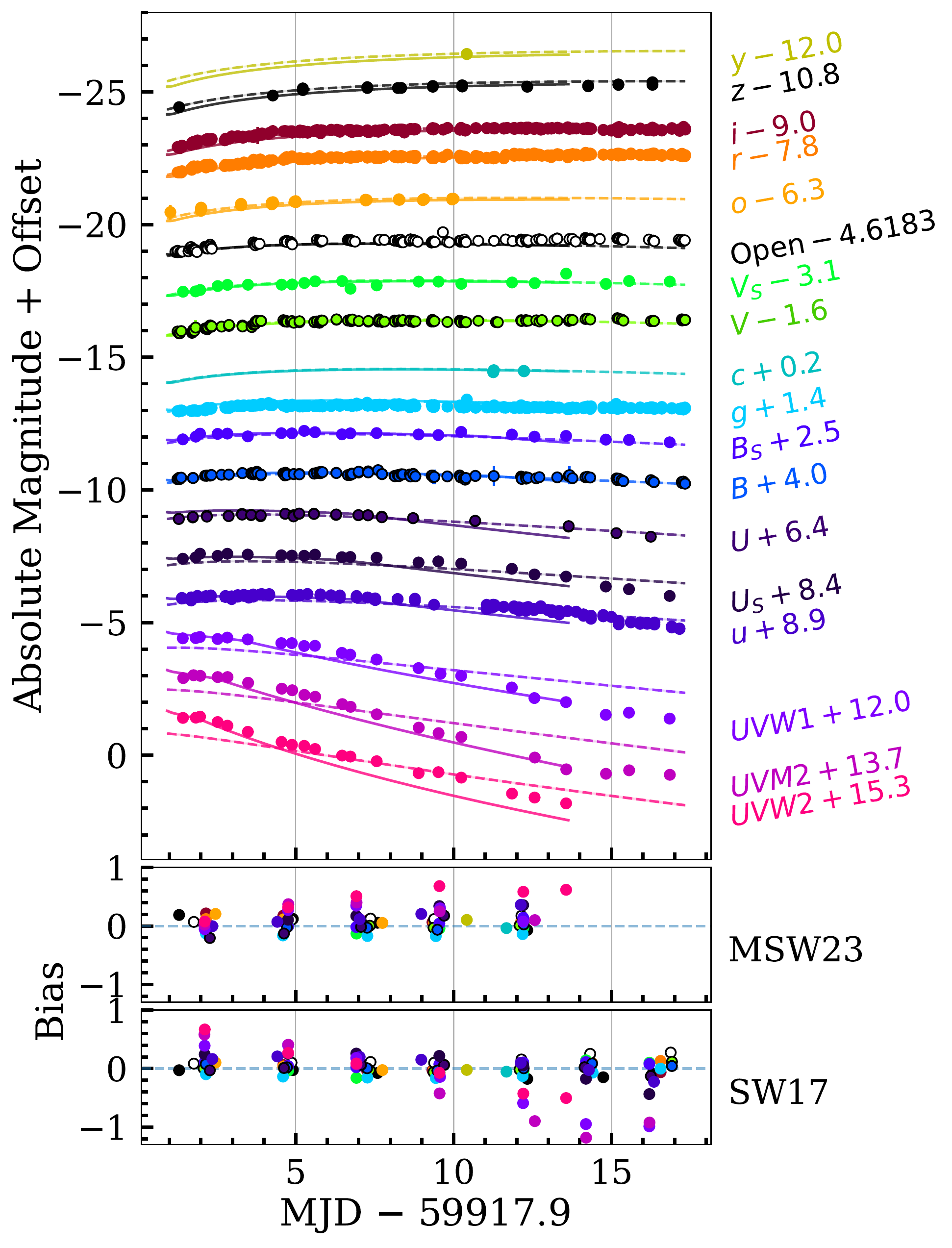}
        \caption{\textit{Upper panel}: Best-fit models from MSW23 (solid lines) and SW17 (dashed lines) compared to the observed early-time light curve. \textit{Lower panels}: Residuals (biases) between the models and the observed data, shown as a function of MJD. Points represent the mean residuals within MJD bins. All models are plotted only within their respective validity time ranges.}
        \label{fig:lcfit}
    \end{figure}

    The phases used in the fit span from MJD 59918.9 to $t_{\rm max}^{\rm model}$, where $t_{\rm max}^{\rm model}$ is defined as the upper validity limit of the models. This threshold corresponds to the epoch when the photospheric temperature drops below $0.7$ eV, at which point the assumption of a fully ionized gas — and thus constant opacity — is no longer valid. These values can be derived from the model parameters -- Eq.~24 from \citet{Sapir_2017} and Eq.~19 from \citet{Morag2023}. We iteratively fit the models decreasing the maximum MJD value for SN~2022acko light curves until we have the $t_{\rm max}^{\rm model}$ above all the data points MJD used in the fit, asserting that the models are valid in every data point used to the fit. We found $t_{\rm max}^{\rm MSW23} = 59931.54\pm0.34$ and $t_{\rm max}^{\rm SW17}=59935\pm2$. Figure~\ref{fig:lcfit} shows the results of our analytical fits compared to the observed data. 
    
    As expected, the MSW23 model exhibits better performance in fitting the UV bands during the very early phases, although both models show overall agreement in the redder bands. This behavior is also evident in the bias dispersion shown in the lower panels of Figure~\ref{fig:lcfit}. Despite their comparable fits to the data, the posterior distributions of the model parameters differ substantially between MSW23 and SW17. Figure~\ref{fig:lcfit} presents the parameter posteriors for both models. 

    \begin{figure*}[ht]
        \centering
        \includegraphics[width=0.495\linewidth]{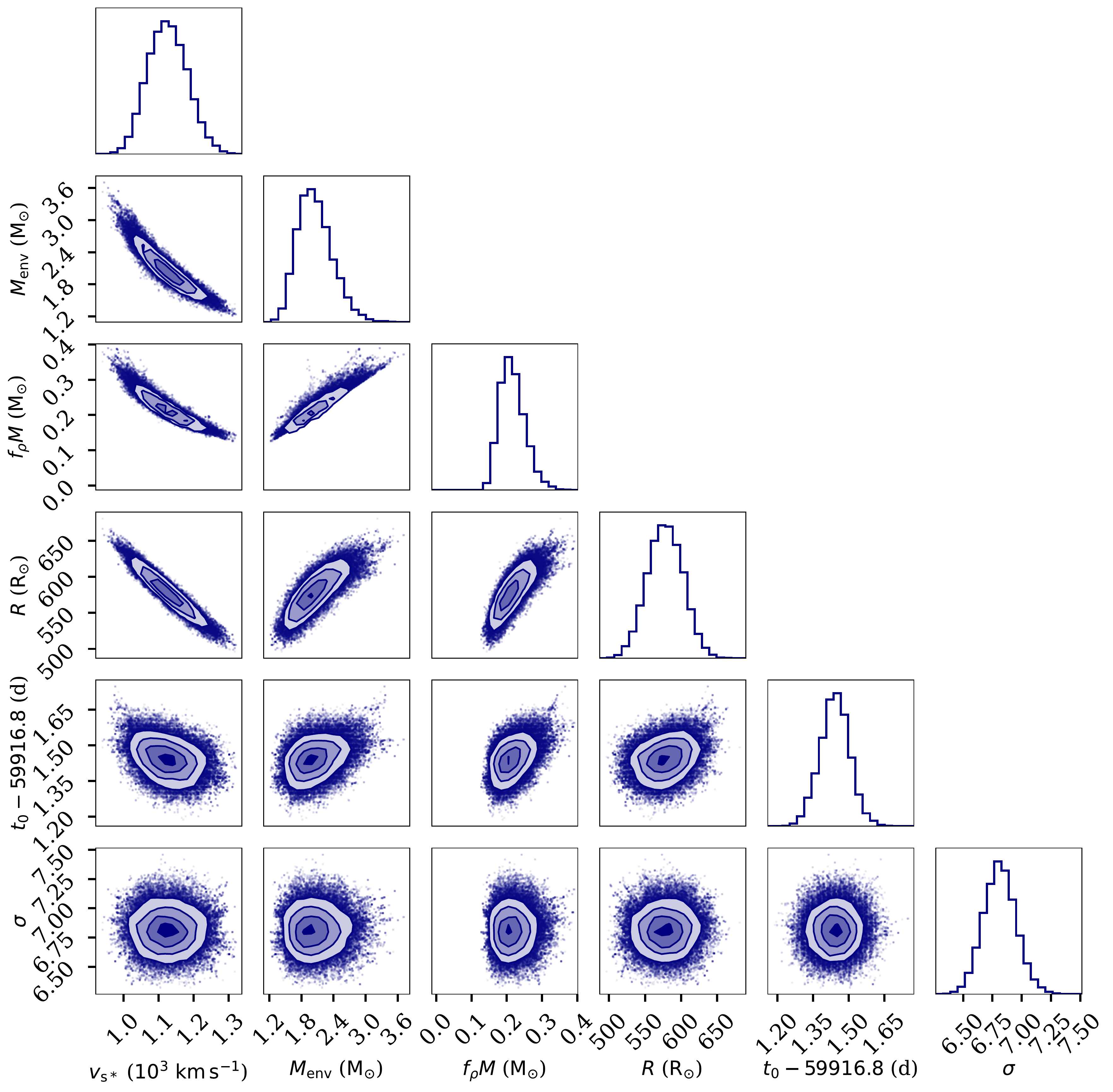}
        \hfill
        \includegraphics[width=0.495\linewidth]{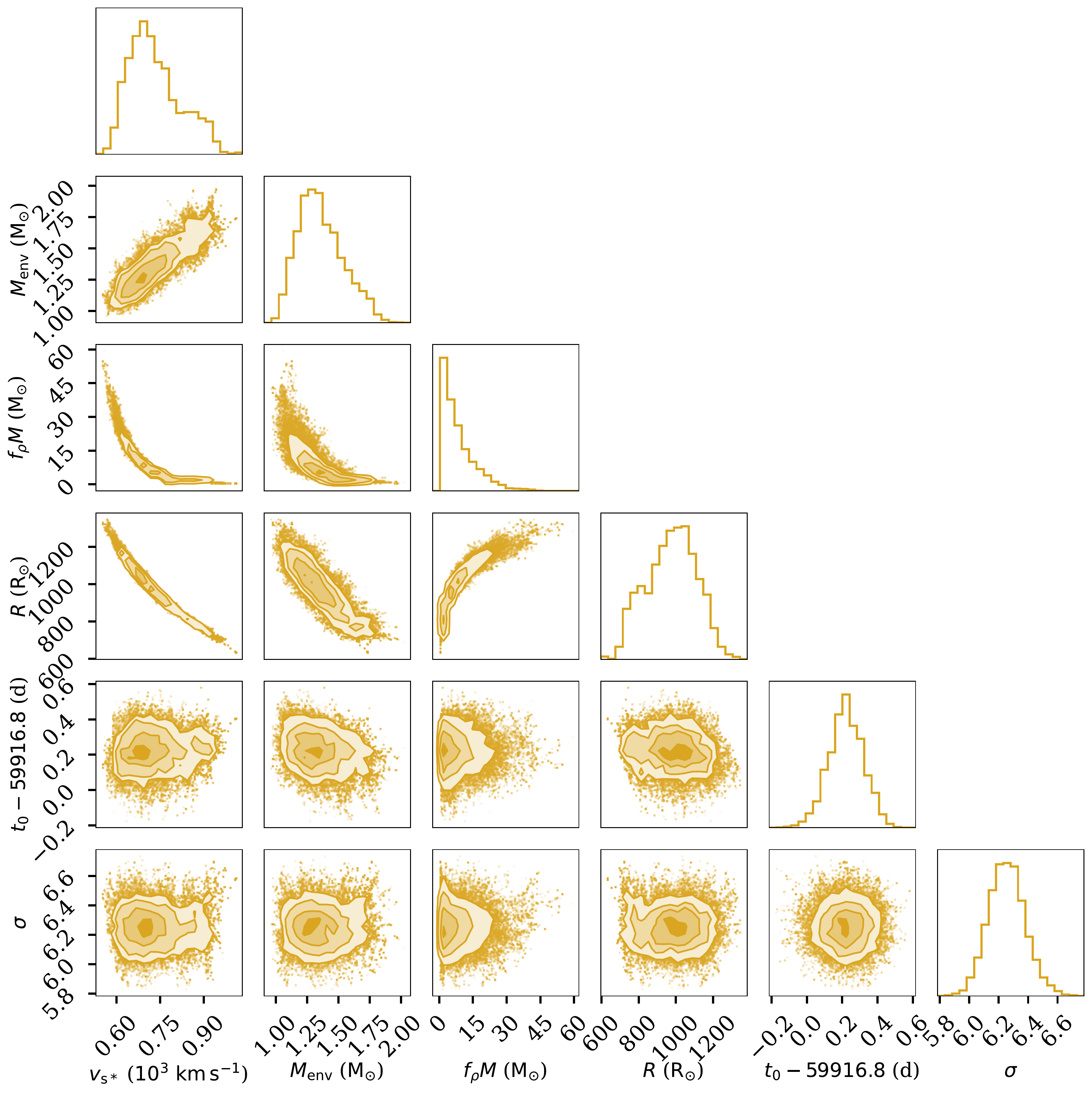}
        \caption{
        Corner plots for the best-fit parameters of SN~2022acko obtained from model MSW23 (left) and model SW17 (right).
        }
        \label{fig:corner_sc3_sc4}
    \end{figure*}

    The SC emission at the early phases is strongly dependent on $R$ and $v_s*$ \citep{Morag2023}, making it essential to have accurate physical measurements of these parameters. Therefore, the disagreement between these parameters among the models cannot be ignored. Additionally, a noticeable discrepancy was found in the $f_{\rho}M$ estimation, in which the values encountered for SW17 are poorly constrained. This is expected for SW17, as \citet{Sapir_2017} shows that the emission is not sensitive to deviations from a polytropic progenitor density profile, resulting in a weak dependence on $f_{\rho}$. Considering this, the MSW23 model, which includes UV suppression at the very early phases of the LC, leads to a better physically described model, and yields more reliable estimates for both $R$ and $v_s*$. In that sense, we chose to base our analysis on the MSW23 estimates of the SN~2022acko progenitor star properties. 
    A similar decision regarding the use of MSW23 for analysis of shock cooling emission is also adopted for other SNe, such as SN 2023ixf, SN 2023axu, and SN 2022jox \citep{Hosseinzadeh_2023, Andrews2024, Shrestha_2024}.

    The best-fit parameters for the MSW23 model, as listed in Table~\ref{tab:shock_fit}, are 
    $v_\mathrm{s*} = 1.12^{+0.06}_{-0.06} \times 10^3\,\mathrm{km\,s^{-1}}$, 
    $M_\mathrm{env} = 2.0^{+0.4}_{-0.3}\Msun$, 
    $f_\rho M = 0.21^{+0.04}_{-0.03}\Msun$, 
    $R = 580^{+20}_{-20}\Rsun$, 
    $t_0 = 59917.44^{+0.06}_{-0.06}\,\mathrm{MJD}$, 
    and $\sigma = 6.8^{+0.1}_{-0.1}$.  
    One notable aspect is that \( f_{\rho M} \) consistently trends toward the lower bound of \(\sim 0.1 \times M_{\rm env}\), which would imply an unusually low ejecta mass. This constraint was manually implemented in the likelihood because, in initial tests, \( f_{\rho M} \) values converged to unphysically small values. Since this trend persists for SN~2022acko, further investigation into the SC modeling is warranted. Moreover, the explosion epoch derived from the SC fit is inconsistent with the last nondetection of SN~2022acko \citep[MJD~59918.17 by ATLAS survey][]{Bostroem_2023}, corresponding to \(\sim 17\)~hours prior to that observation.

\begin{figure}
        \centering
        \includegraphics[width=\linewidth]{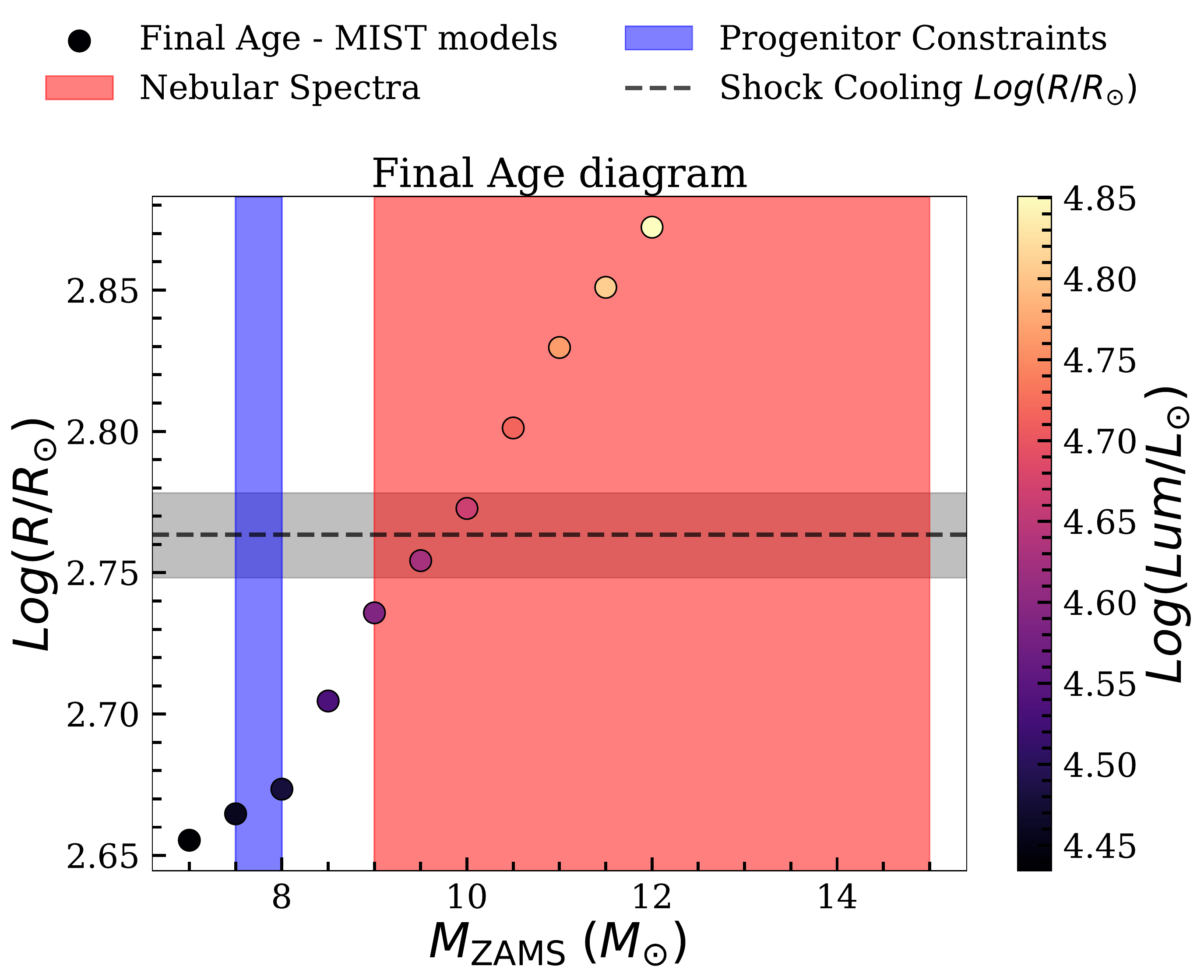}
        \caption{The terminal radius and luminosity from stars evolution tracks in function of the stars initial masses}
        \label{fig:logRlogT}
    \end{figure}

    The radius of \(R = 580^{+20}_{-20}\Rsun\) found by MSW23 for the progenitor is compatible with the expected range of \(100\)–\(1000\Rsun\) \citep{Levesque2020}. To estimate the progenitor mass prior to explosion, we compared this radius from our SC modeling with the terminal radii of stars at their final evolutionary stage, as given by the MESA Isochrones \& Stellar Tracks \citep{mist}. Assuming a metallicity \(\mathrm{[Fe/H]} = -0.22\), corresponding to the metallicity derived for the disk region of NGC\,1300 \citep{rosado_2020}, Figure~\ref{fig:logRlogT} shows the terminal radius and luminosity of stars at their final evolutionary stage for different values of \(M_{\rm ZAMS}\), along with the progenitor mass estimates derived in this work. This analysis suggests a progenitor \(M_{\rm ZAMS}\) of \(9\)–\(10\Msun\), consistent with typical values for RSGs \citep{davies_2018}.

\subsection{Bolometric light curve and nickel mass}
\label{subsec:bolometric}

Our dataset includes coverage in 17 bands, with well-sampled optical/UV data up to approximately 40 days post-explosion, and optical data with good cadence extending to the later phases (up to approximately 350 days). We take advantage of this extensive coverage to reconstruct the bolometric luminosity of SN~2022acko. We used the \texttt{extrabol} software \citep{thornton2024extrabol}, a \texttt{python}-based package that employs Gaussian Process Regression (GPR) to interpolate the multi-band light curves.

For the first 130 days, we estimated the bolometric luminosity by linearly interpolating the in-band light curves over time and fitting a blackbody model to each epoch. After this phase, blackbody fits become inappropriate, as hydrogen recombination fades and radioactive decay dominates the emission. Therefore, for the later epochs, we set \texttt{extrabol} to continue interpolating the light curves but switch to fitting nebular models to estimate the bolometric luminosity.

To convert our late-time photometry into bolometric luminosities, we require an estimate of the spectral energy distribution (SED) at each epoch. We use theoretical nebular spectra from \citet{jerkstrand_9m, jerkstrend_12m} and \citet{dessart2021} as SED templates. Specifically, the \texttt{extrabol} routine compares the observed photometry to each model spectrum in the $9$--$29~\rm \Msun$ grid, scaling the spectra to match the photometric fluxes. The model providing the best overall match to the photometry is then adopted as the reference SED shape, which is kept fixed while rescaling its normalization to match each photometric epoch. This approach is independent of the direct comparison between the observed nebular spectra and models described in Section~\ref{sec:nebular_spec}.

In this way, the later epochs are reduced to a single-parameter fit to the photometry with a similar reduction in uncertainties on the overall luminosity as observed in Figure~\ref{fig:lcfit}.  In order to account for the transition between the optically thick and thin photosphere -- modeled as blackbody and nebular spectral energy distributions, respectively-- we further incorporate an apodization timescale of $t_\Delta=30$~days where both models are fit simultaneously with a weighting that transitions from a purely blackbody to purely nebular spectrum. The transition is centered at  $t_s=130$~days, marking the epoch when the dominant emission shifts from blackbody to nebular. The weighting is defined as a \textit{Fermi-Dirac};

\begin{equation}
    w = \frac{1}{1+e^{-(t-t_s)/\Delta t}} \,,
\end{equation}

\noindent smoothly varying from 0 to 1, such that the total bolometric luminosity is given by

\begin{equation}
    L(t) = (1-w)L_{BB}(t) + wL_{neb}(t) .
\end{equation}

\begin{figure}
    \centering
    \includegraphics[width=1\linewidth]{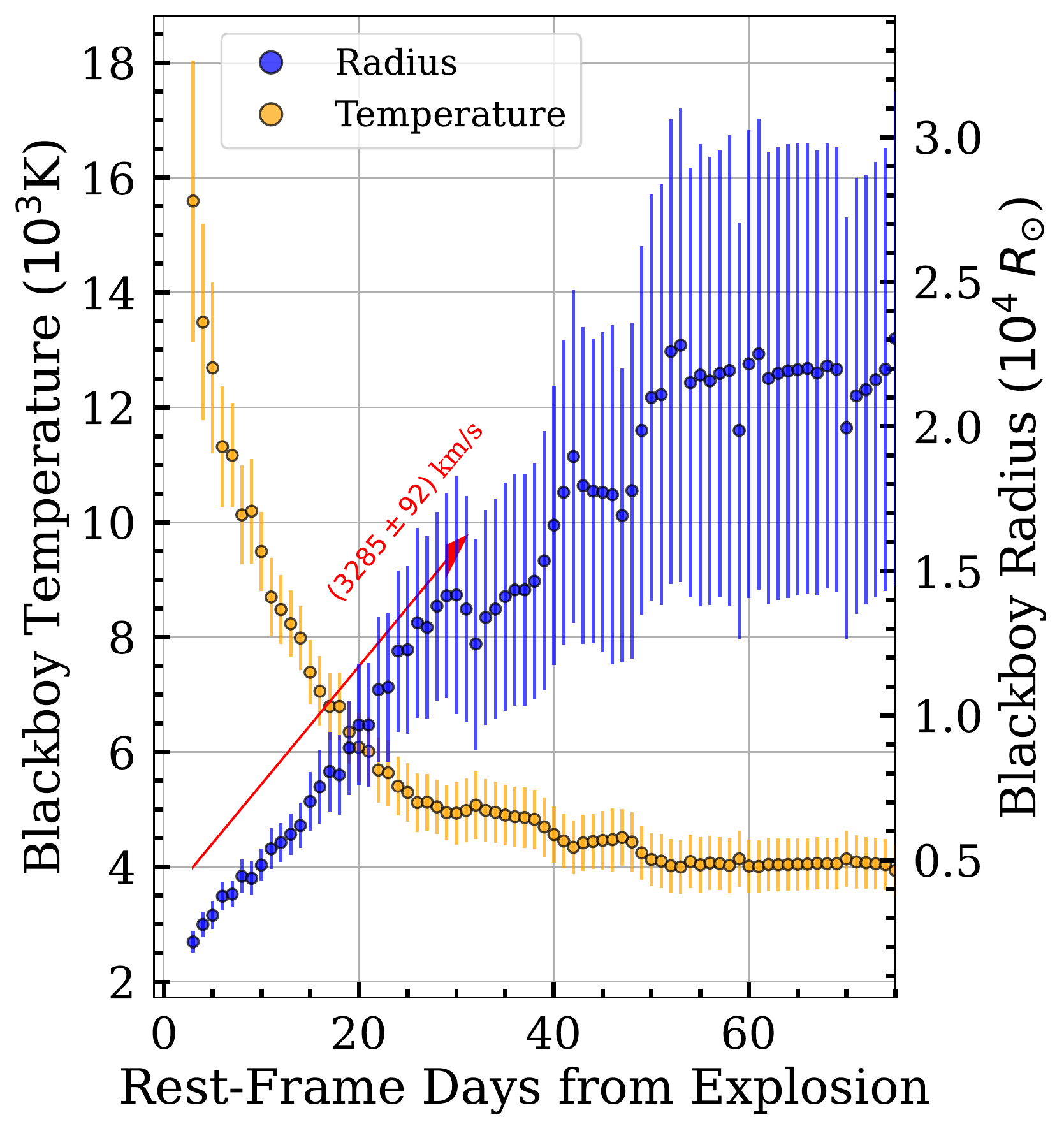}
    \caption{The radius (blue dots; right axis) and temperature (orange dots; left axis) evolution derived from the bolometric lightcurve fitting of SN2022acko. } 
    \label{fig:RTevol}
\end{figure}

Figure~\ref{fig:RTevol} shows the evolution of the blackbody radius and temperature up to 70 days after the explosion, as derived in Section~\ref{subsec:sc}. From this result, we determined a photospheric velocity of 3285~km\,s$^{-1}$ by assuming a linear evolution of the radius up to 30 days post-explosion. This value is lower than the typical expansion velocities observed for \sneiip\ \citep{Bose2014}, but is comparable to the mean velocities measured for SN~2009md and SN~2008in -- both \sneiip\ events analyzed in the same study.

\begin{figure}
    \centering
    \includegraphics[width=1\linewidth]{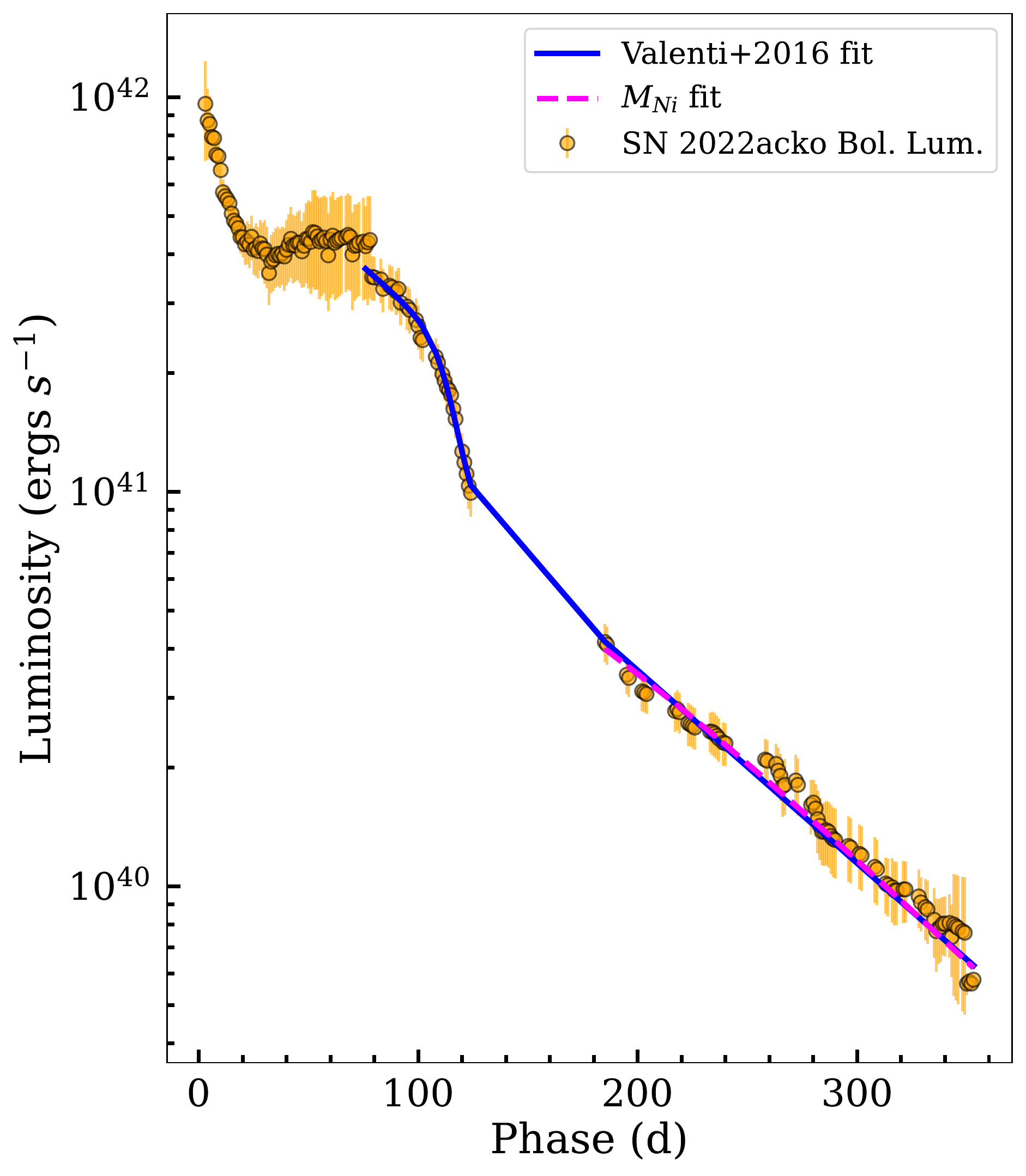}
    \caption{The bolometric luminosity (orange circles) derived using extrabol for SN~2022acko. We also displays the parametric light curve model from \citealt{Valenti2016} fitted in our results (solid blue line) for $MJD>25$. } 
    \label{fig:bololight}
\end{figure}

We also present the derived bolometric light curve in Figure~\ref{fig:bololight}. Following the formalism outlined by \citet{Valenti2016} --see Equation 2 in their work--, we determined a plateau duration for SN~2022acko of $t_{\rm PT} = 117$ days, which is slightly above the mean of the distribution observed in their SNe sample ($112\pm 15$~days as shown Figure~9 in their work). The extended plateau suggests that the progenitor of SN~2022acko possessed a massive hydrogen-rich envelope. This is consistent with the independent mass constrains from the shock cooling modeling, which also favor a high envelope mass.

From the same fit, we inferred a luminosity decline rate during the nebular phase of approximately $1.22 \rm \,mag/(100\days)$. This value is deviating about $20\%$ from the expected decay rate of $\cofs$, assuming complete gamma-ray trapping \citep[$0.98 \rm mag/(100 \; days)$; see][]{Woosley1989}. This suggests a significant impact of gamma-ray leakage on SN~2022acko's radioactive emission phases, which we explore below.

In order to derive the total $\nifs$ mass produced in SN~2022acko, we follow the methods outlined by \citet{Hamuy2003}, \citet{kilpatrick2023type}, and \citet{Tinyanont2021}. We assume that the luminosity 117 days post-explosion, which is after the decline from plateau based on our \citet{Valenti2016} modeling, is primarily dominated by the decay of $\cofs \rightarrow {}^{56}\rm Fe$. We further modify these models to account for gamma-ray leakage, fitting an analytic model to the bolometric luminosity where the total $\nifs$ mass follows

\begin{equation}
\label{eq:nickelmass}
    M_{\rm Ni} = \frac{L(t)}{\epsilon_{\rm Co}}
    \frac{\lambda_{\rm Co} - \lambda_{\rm Ni}}{\lambda_{\rm Ni}}
    \left(
        e^{-\lambda_{\rm Ni} t}-e^{-\lambda_{\rm Co} t}
    \right)^{-1}
   f_{\rm leak}^{-1}
\end{equation}

\noindent where $L(t)$ is the bolometric luminosity at rest-time $t$ after the explosion, $\epsilon_{\rm Co} = 6.8 \times 10^{9}$~ergs~s$^{-1}$~g$^{-1}$ is the heating rate from the decay of $\cofs$, $\lambda_{\rm Co} = 1/111.4 \ \rm d^{-1}$ and $\lambda_{\rm Ni} = 1/8.8 \ \rm d^{-1}$ are the radioactive decay timescales for $\cofs$ and $\nifs$, respectively, and $f_{\rm leak} = 1 - 0.965 \times \exp(-(t/t_1)^2)$ is a factor accounting for incomplete trapping of gamma-rays emitted by the decaying $\cofs$ and $\nifs$.  In this equation, $t_1$ is a parameter determining the timescale for the optical depth of gamma-rays to decrease to 1 \citep[see][]{Sollerman1998}. We found $t_1 = 396 \pm 7 \rm d$.  In theory, this value is correlated with the total ejecta mass \citep[e.g.,][]{Sollerman1998}, with higher ejecta mass correlating with a longer transparency timescale.  Comparing with events such as SN~2020jfo that have well-sampled light curves \citep{Sollerman2021,kilpatrick2023type}, we find a significantly longer timescale.

By fitting Equation~\ref{eq:nickelmass}  to our bolometric light curve at rest-frame times greater than $\sim 170$ days post-explosion (post seasonal break), we derived the total $\nifs$ mass to be $M_{\rm Ni} = 0.014 \pm 0.004\Msun$. The primary source of uncertainty in this measurement arises from the systematic uncertainty in the luminosity distance to the host galaxy of SN~2022acko. The uncertainty in $M_{\rm Ni}$ was computed assuming a luminosity distance of $d_L = 19.0 \pm 2.9$~Mpc for NGC 1300. This value follows \citet{Bostroem_2023}, who adopted the distance estimate from the PHANGS survey \citep{Anand2020}, where it was derived using the Numerical Action Method \citep[NAM, ][]{Shaya_2017,Kourkchi_2020}. 

However, a broader range of distance estimates to NGC\,1300 would result in larger systematic uncertainties on $\nifs$ as in \citet{Scheuermann2022} who derive an upper limit of $25.77^{+0.90}_{-1.42} \,\mathrm{Mpc}$. 
They also compiled previous distance estimates for NGC 1300 (including derivations from NAM; see their Fig. A3), showing that even among the most precise measurements, values range from approximately 14-26~Mpc. This range would imply in a corresponding $M_{\rm Ni}$ for SN 2022acko between $0.007-0.022$~$\Msun$. 

Nevertheless, the of $M_{\rm Ni}$ derived SN 2022acko in this work appears to lie within the typical range observed for \sneiip\ \citep[i.e., those in][]{Valenti2016,Muller2017,Martinez2022}. It is consistent with the range of values reported for a sample of 16 SNe in \citet{Valenti2016} ($\sim 0.006-0.18 \smasm$ as shown Figure~10 ).

\begin{table}[h]
    \centering
    \caption{Values extracted from fitting the SN~2022acko bolometric light curve.}\label{tab:lc_fit}
    \begin{tabular}{|c|c|}  
        \hline
    $\bm{A0}$ & $-0.41\pm0.02$\\\hline
    
    $\bm{T_{PT}}$ \textbf{(d)} & $117.7\pm0.6$\\\hline
    
    $\bm{W0}$ \textbf{(d$^{-1}$)} & $5.3\pm0.7$\\\hline
    
    $\bm{P0}$ \textbf{(d$^{-1}$)} & $-0.00489\pm0.00007$\\\hline
    
    $\bm{M0}$ & $41.52\pm0.02$\\\hline
    
    $\bm{M_{\rm Ni}}$ \textbf{($\text{\textbf{M}}_{\odot}$)} & $0.0145\pm0.0003$\\\hline
    
    $\bm{t_1}$ \textbf{(d)} & $384\pm11$\\\hline

    \end{tabular}
\end{table}

    
    
    
    
    
    


\subsection{Nebular spectroscopy analysis}
\label{sec:nebular_spec}

\begin{figure}[ht]
    \centering
    \includegraphics[width=.9\linewidth]{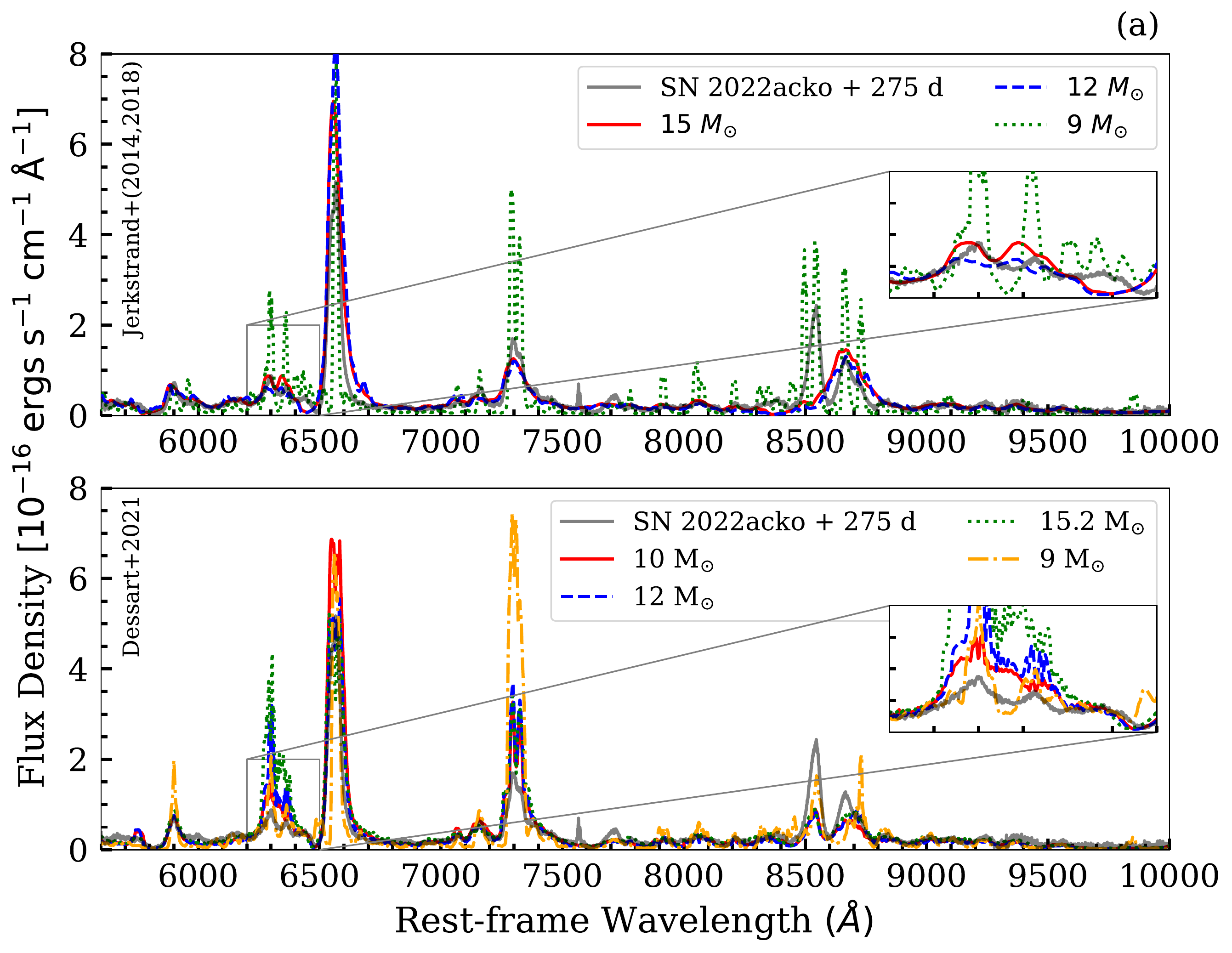} 
    
    \vspace{.05cm}
    
    \includegraphics[width=.9\linewidth]{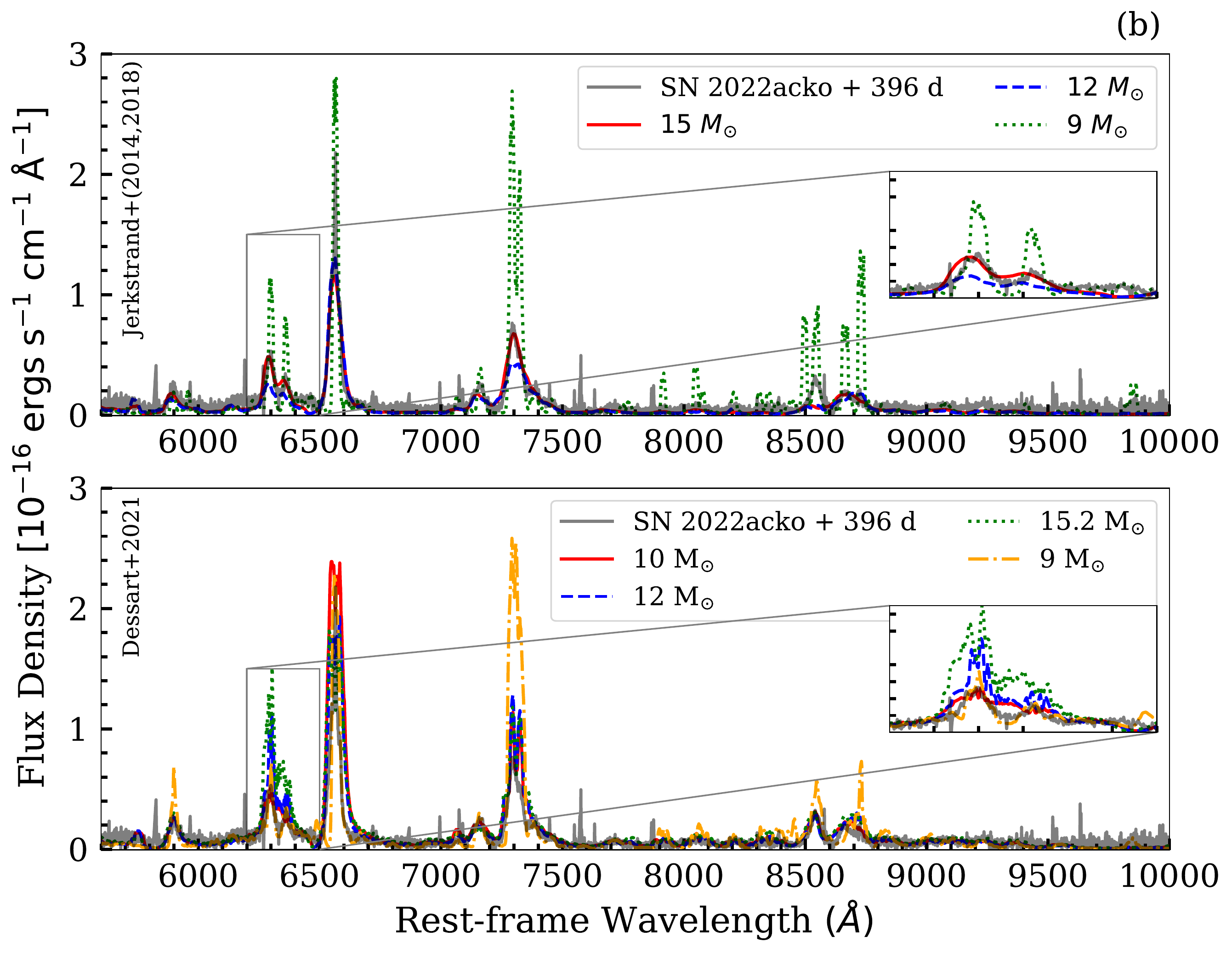} 
    
    \vspace{.05cm}
    
    \includegraphics[width=.9\linewidth]{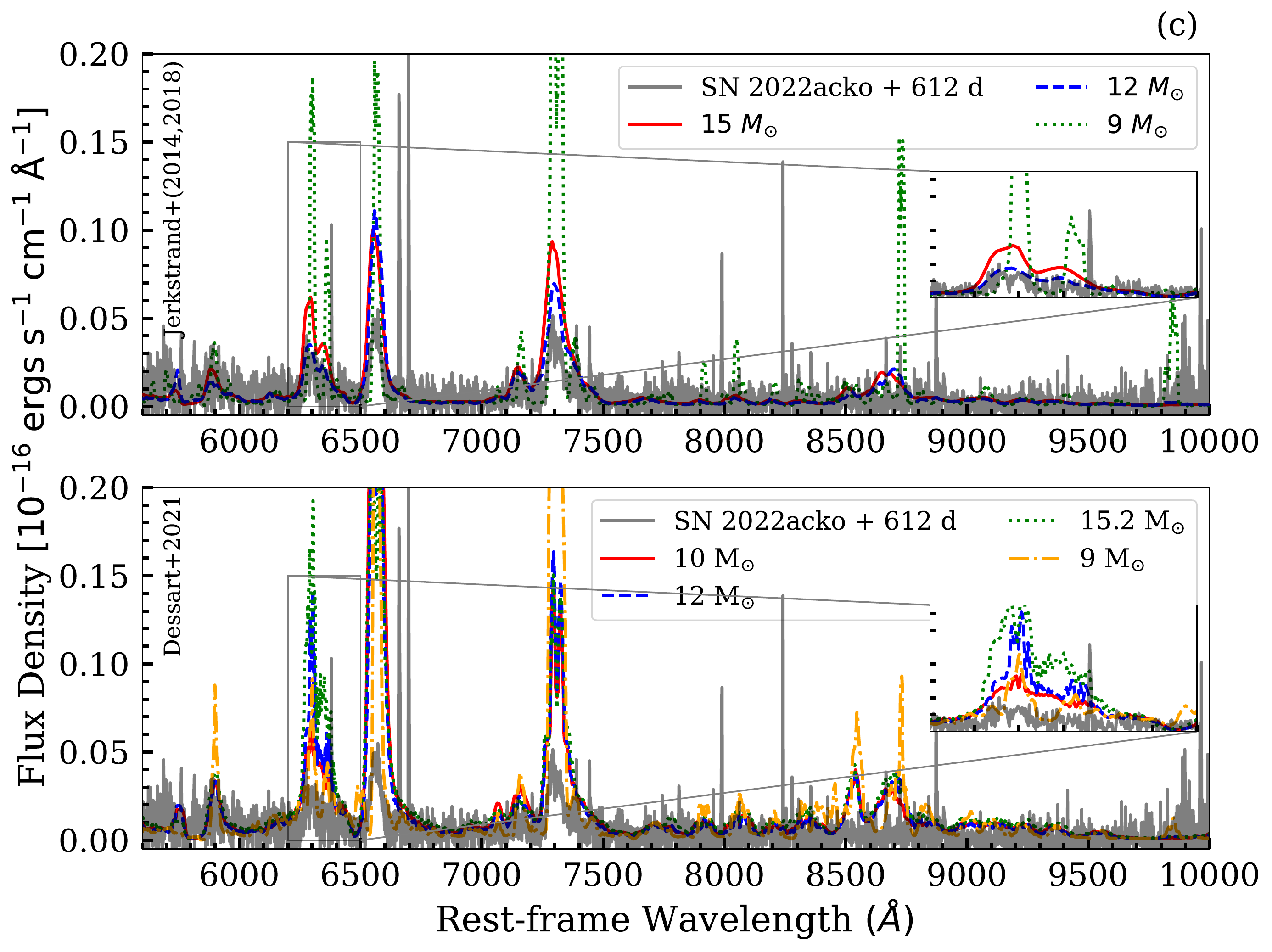} 

    \caption{SN 2022acko Keck spectra (grey lines) from September 2023 (a), January 2024 (b), and August 2024 (c)  at 275, 396, and 612 days from rest frame explosion date, respectively. We compare both spectra with nebular models from \citet{jerkstrand_9m} and \citep{jerkstrend_12m} (upper pannels), and \citet{dessart2021} (lower pannels).  Within each panel, we show an inset zoomed in on the [O\textsc{i}] $\lambda\lambda$6300, 6364 lines, which are the main features used to match to models of \snii\ nebular spectra.}
    \label{fig:nebular_spectra}
\end{figure}

The nebular phase of SNe begins when the ejecta become optically thin \citep{Ferrari2024, kilpatrick2023type, Tinyanont2021}. During this phase, the spectra are dominated by emission lines from the synthesized elements present in the nebular interior. Among these, the strength of the oxygen lines ([OI] at 6300 and 6364 \AA) is particularly sensitive to the progenitor’s core oxygen mass, making it a key diagnostic for estimating the progenitor star initial mass \citep{Jerkstrand2017}. 
Modeling these nebular spectra is possible by accounting for factors such as the nebular composition, velocity fields, non-thermal processes, and Non-Local Thermodynamic Equilibrium (NLTE) gas states, among others \citep{Jerkstrand2017}. Here, we present a comparison between synthetic nebular models for \sneii\ and the observed spectra of SN~2022acko, described in Section~\ref{sec:observation-spec}, to estimate the initial mass of the SN~2022acko progenitor star.  

We used two different sets of nebular models in this analysis. Both sets are based on NLTE radiative transfer and nucleosynthetic yields for core-collapse SNe. The first set encompasses progenitor stars with initial masses of 9$\rm \Msun$ \citep{jerkstrand_9m} and 12-29$\rm \Msun$ \citep{jerkstrend_12m}, assuming nebular spectra of different ages. The second set consists of a grid of models for progenitor stars with initial masses between 9-29$\rm \Msun$, corresponding to one-year-old \sneii\ \citep{dessart2021}.

All considered models were scaled by the ratio between the total $M_{\rm Ni}$ derived for SN~2022acko (Section~\ref{subsec:bolometric}) and the $M_{0,\rm Ni}$ value assumed for generating each of the synthetic spectra. Furthermore, we scaled the modeled spectra to match the distance of SN~2022acko (19.0 Mpc, see Section~\ref{subsec:bolometric}). Finally, we accounted for the difference between the observation dates of the SN~2022acko spectra and the ages of the SNe used to generate the synthetic spectra, $\delta t$, by scaling the fluxes by $\exp{-\lambda_{\rm Co}\delta t}$. At the end, the modeled fluxes were rescaled by the factor $s_0$, given by

\begin{equation}
    s_0 = \frac{M_{\rm Ni}}{M_{\rm Ni,0}}
    \times
    \left(\frac{d_{L,0}}{d_{L}}\right)^2
    \times
    e^{-\lambda_{\rm Co}\delta t} \, ,
\end{equation}

\noindent in which $d_{L,0}$ is the luminosity distance assumed for generating the synthetic spectra. The observed spectra were calibrated for the expected flux of the SN at each observation date as implied by the light curves. We used \texttt{pysynphot} \citep{pysynphot} to generate synthetic photometry from the spectra in the $r$ and $i$ bands for the 275-day-old spectrum, in the $i$ band only for the 396-day-old spectrum, and in the $z$ band for the 612-day-old spectrum. The fluxes were then scaled so that the synthetic photometry matched the observations.

The photometry dates, $t_{phot}$, do not exactly match the spectroscopic dates, $t_{spec}$. Therefore, after the rescaling, we further adjusted the flux by multiplying it by the factor $\exp{-\delta t \lambda_{\rm Co}}$, where $\delta t_{cal} = t_{spec} - t_{phot}$. We emphasize that the photometry used for rescaling was chosen to be as close as possible to $t_{spec}$ for each spectrum. The values of $\delta t_{cal}$ were -2, +42, and +260, corresponding to the spectra dates in ascending order.

Figure~\ref{fig:nebular_spectra} shows the flux-calibrated observed nebular spectra from Keck, alongside the models of \citet{jerkstrand_9m, jerkstrend_12m} and \citet{dessart2021}. Notably, for the two youngest spectra (panels \textit{a} and \textit{b}), there is a disagreement between the progenitor star mass suggested by the two models that best fit the observations. For the earliest spectra (275 d), the comparison with \citet{jerkstrand_9m, jerkstrend_12m} models suggests that the progenitor star of SN~2022acko is more likely a 12 $\Msun$ star. On the other hand, there is no significant agreement between the spectra with any of \citet{dessart2021} models in the oxygen doublet region, except for small intersection with the 9 $\Msun$ model. However, if we consider that this disagreement may rely in the calibration uncertainties, and then focus in comparing the shape of the spectra in this region, the structure of the 10 $\Msun$ is more similar to the measured spectra. 

When analyzing the 396 d old spectra, we have also a disagreement, but now the best fit for the \citet{jerkstrand_9m, jerkstrend_12m} models corresponds to a 15 \smast progenitor, while the the spectra show better agreement with a 9 $\Msun$ progenitor star when compared to the \citet{dessart2021} models.

It is worth noting that both models suggest a progenitor star mass of $>9;\Msun$. However, these results are in disagreement with the estimates of the progenitor star mass derived from the direct observations (Section~\ref{sec:progenitor}).


The analysis of the oldest spectrum presents two main caveats. First, we did not have SN~2022acko photometry available near the epoch of the spectrum, so the flux calibration may not accurately reflect the true values. Despite this limitation, it is encouraging to note that pronounced emission lines, including those from [O~\textsc{i}] $\lambda$6303, H$\alpha$, and [Ca~\textsc{ii}] $\lambda\lambda$7922, 7324, are still visible. 

The second caveat is that this spectrum corresponds to one of the latest phases ever observed for a \sniip\, a regime where the validity of existing nebular models has not been thoroughly tested.  Although the spectrum is significantly lower in signal-to-noise than the earlier phases, we detect the expected spectral features that are associated with nebular \sniip\ in particular the [O~\textsc{i}] feature. However, the nebular models we use only model the evolution of \sneiip\ up to 500 days post-explosion in the case of the \citet{jerkstrand_9m, jerkstrend_12m} models, and around 350 days post-explosion for the \citet{dessart2021} models. 

Although this results may be strongly affected by the aforementioned caveats, for the late-time spectrum, the best agreement — in terms of the [O\textsc{i}] doublet — is found with the $12\Msun$ progenitor model from \citet{jerkstrand_9m, jerkstrend_12m}.

The trustworthiness of the comparison between the nebular spectra and theoretical models is compromised by significant uncertainties in flux calibration and by systematic errors introduced when rescaling the models to account for differences in nickel mass and luminosity distance. These limitations likely contribute substantially to the discrepancies discussed above, once the scaling factor alone introduces a relative uncertainty of approximately 45\%.

\section{Discussion}
\label{sec:discussion}
The progenitor mass estimates from pre-explosion imaging, shock cooling modeling, and nebular spectroscopy are summarized in Table~\ref{tab:mass_estimates}. These estimates rely on three different natures of data, and lead to a very broad range of reliable constrains for the progenitor mass.

\begin{table}[ht]
\centering
\caption{Summary of progenitor mass estimates for SN~2022acko.}
\label{tab:mass_estimates}
\begin{tabular}{lcc}
\toprule
\textbf{Analysis} & \textbf{Mass Estimate ($\Msun$)} & \textbf{Section}\\
\midrule
Pre-explosion imaging & $7.5 - 8.0$ & \ref{sec:progenitor_sed} \\
Shock cooling modeling & $9-10$ & \ref{subsec:sc} \\
Nebular spectroscopy & $9-15$ & \ref{sec:nebular_spec} \\
\bottomrule
\end{tabular}
\end{table}

Using early-phase photometry, we applied the shock cooling emission models from \citet{Morag2023} to constrain the progenitor star’s radius to $\sim580\Rsun$. By comparing this estimate with MESA evolutionary tracks, we inferred a progenitor mass of $9-10\Msun$ for SN~2022acko. These findings are consistent with typical values of progenitor mass for \sneii, and are also consistent with the lower estimates from the nebular spectroscopy modeling.

Although SC modeling results are compatible with a higher mass progenitor, these models involve several approximations, and their systematics are not yet fully understood. In the specific case of SN~2022acko, the SC analysis suggests that the first detection occurred approximately 1.5 days after the estimated date of explosion. Since SC-based estimates of radius—and by extension, mass—depend critically on early-time data, we performed a rough test to estimate the impact of a delayed first detection.

As test case, we used SN~2021yja \citep{Hosseinzadeh2022_2021yja}, a well-observed event with low CSM density and a discovery only $\sim$5.4~hours post-explosion. Using the MSW23 model, we first fit the SC emission using the initial 9 days of data, as done in \citet{Hosseinzadeh2022_2021yja}. We then repeated the fit, excluding the first 1.5 days to simulate the data gap similar to SN2022acko. The resulting posteriors (Figure~\ref{fig:2021yjacorner}) indicate that omitting early-time data can lead to a significantly larger (about $\sim100\Rsun$) inferred progenitor radius. While this exercise is approximate, it suggests that the SC-derived radii--and thus masses—-for SN~2022acko could be overestimated if early emission is missing.

We also compared the nebular-phase spectra of SN~2022acko with models from \citet{jerkstrand_9m, jerkstrend_12m} and \citet{dessart2021}, and found that none of the three independent spectra yielded consistent progenitor mass estimates across the different models. The variation in outcomes from these visual comparisons results in a poorly constrained progenitor mass range spanning approximately $9-15\Msun$. This highlights a tension between two distinct mass regimes: a higher-mass scenario ($9-15\Msun$, higher relative to the very low mass estimate from progenitor observations) supported by shock-cooling modeling and nebular spectroscopy, and a lower-mass scenario ($\sim7.5$--$8.0\Msun$) derived from direct detection of the progenitor through RSG SED fitting. Further contributing to this discrepancy, the estimated $\nifs$ mass of $0.014\Msun$ synthesized in the explosion--despite the intrinsic scatter in its correlation with progenitor mass-—also favors a more massive progenitor.

A possible explanation for the disagreement in progenitor star mass estimates is significant envelope loss during the lifetime of SN~2022acko progenitor star. In such a scenario, the progenitor could have had a higher initial mass, reflected in the nebular-phase spectra due to the strong correlation between progenitor mass and synthesized oxygen. While the observational features closer to the explosion would appear consistent with a lower-mass star, after substantial envelope stripping. This scenario would be further supported if shock-cooling (SC) modeling tends to overestimate the progenitor radius, since such a close-to-explosion measurement would be consistent with a low progenitor mass. However, to adopt this assumption with confidence, additional studies are needed to better characterize and quantify the systematic uncertainties and biases inherent in SC models.

If this significant mass loss were driven primarily by stellar winds or episodic outbursts \citep[see review in][]{Smith2014}, we would expect clear evidence of CSM surrounding the progenitor, such as a secondary luminosity peak due to SC UV re-emission in the early light curve \citep{Morag2023, Waxman_2017}. Even if CSM were present, however distant from the star, we would anticipate a signature on timescales of years assuming the mass loss occurred in the final 1000~yr prior to core collapse \citep[i.e., during carbon burning or later;][]{Limongi24}. The shock would heat this material, resulting in an increase in luminosity from thermal radiation. However, apart from fluctuations comparable to the plotted uncertainties, Figure \ref{fig:bololight} shows a nearly steady decay in luminosity for the later phases, which does not indicates the presence of CSM at extended radii from the progenitor star \citep[as opposed to the confined CSM in][]{Bostroem_2023}. At this stage, mass transfer to a binary star companion emerges as an alternative explanation, which we explore below.

Binary systems can reach mass transfer rates of $\sim 0.01\Msun\,\rm yr^{-1}$ \citep{Blagorodnova2021}, potentially explaining the mass loss with absence of strong CSM effects in our observations. While no direct evidence of a binary companion at the site of SN~2022acko is available (see Section \ref{sec:progenitor}), we explored this hypothesis using Binary Population and Spectral Synthesis \citep[BPASS, ][]{Byrne2022} models.

To constrain the progenitor system, we examined BPASS evolution models for NGC 1300's metallicity \citep[$\rm Fe/H= -0.22$; ][]{rosado_2020}. We compared the final predicted magnitudes of these models in the F814W and F160W filters with the progenitor photometry from Section \ref{sec:observations}, using a 3$\sigma$ tolerance. Among the models meeting the photometric constraints, none with a terminal mass $\gtrsim 7\Msun$ were found to simultaneously match our progenitor constrains(Table \ref{tab:mass_estimates}). This suggests that no reliable high-mass binary evolution model in the BPASS grid adequately reproduces the observed properties of SN,2022acko.

While binary mass transfer remains a plausible mechanism, further evidence--such as late-time imaging or detailed modeling--will be necessary to confirm its role in shaping SN2022acko’s progenitor evolution and explosion properties.


\section{Conclusions}
\label{sec:conclusions}
We have presented an extended optical and UV light curve for SN~2022acko, covering approximately 350 days. The final $\sim$200 days of observations represent a novel compilation, including original data from the STEP follow-up program. Our dataset also includes pre-explosion imaging from {\it HST} at the SN~2022acko site, as well as three epochs of Keck spectroscopy obtained at 275, 396, and 612 days after detection. We used this dataset to derive independent constraints on the progenitor star’s properties, with a particular focus on constraining the star's initial mass.

Using \textit{Spitzer} and \textit{HST} pre-explosion imaging of the SN~2022acko site --specifically the F814W and F160W bands from \textit{HST}, along with photometric upper limits from other filters-- we fit a red supergiant (RSG) spectral energy distribution (SED) and found better agreement with models corresponding to a progenitor mass of approximately 7.5 solar masses.

Applying SC emission models from \citet{Morag2023} to the SN eraly-phase photometry, we constrained a progenitor radii of $\sim580\Rsun$. Assuming a typical terminal effective temperature for RSGs of $\sim$3500\,K, this corresponds to a progenitor luminosity of $\log(L/\Lsun) \approx 4.66$. We compared these estimates with MESA evolutionary tracks, and found this radii consistent with progenitors star with mass within $9-10\Msun$. 

We compared the nebular-phase spectra of SN~2022acko with models from \citet{jerkstrand_9m, jerkstrend_12m} and \citet{dessart2021}, finding an inconsistency with the low-mass progenitor scenario suggested by the pre-explosion analysis. Instead, the spectral features show better agreement with models corresponding to progenitor masses in the range of $9-15\Msun$. Additionally, the estimated $\nifs$ mass synthesized in the explosion also supports a more massive progenitor.


To explain the discrepancy among the progenitor mass estimates, we considered the possibility that SN~2022acko originated from a binary system and had undergone significant mass loss through interaction with a companion. However, upon analyzing binary evolution models from the BPASS database, we found no scenario that simultaneously satisfies both the mass constraints and the pre-explosion {\it HST} observations at the SN~2022acko site.


The systematic uncertainties --approximately 45\%-- in the flux calibration of the nebular spectra limit the reliability of the higher progenitor mass estimates. As a result, the tension between mass estimates tends to favor the lower values inferred from shock-cooling modeling and pre-explosion photometry. These constraints suggest a progenitor mass in the range of $7.5$–$10$,\smast, but further investigation is needed, particularly regarding the systematics of shock-cooling models, which appear to be highly sensitive to how early the post-explosion data were obtained.


Nevertheless, the lower mass estimates are also subject to notable limitations. In addition to relying on upper limits, the pre-explosion SED fitting is based on only two photometric detections (F814W and F160W), which might not be sufficient to robustly constrain the full spectral energy distribution. Similarly, the shock-cooling models are based on simplifying assumptions that may not fully reflect the physical conditions of SN~2022acko. Indeed, previous studies employing these models have encountered difficulties in accurately reproducing early-time light curves, especially in the UV regime \citep{2021gmj, 2024ggi, 2022jox, 2023axu}.

This work presents a comprehensive analysis encompassing data from pre-explosion imaging, early-time photometry, and late-time nebular spectroscopy. The tension between different progenitor mass estimates does not appear to stem primarily from a lack of observational data, but rather from intrinsic uncertainties in flux calibration and the limitations of the current models. A more detailed and physically realistic modeling, as well as a deeper investigation on the sources of systematic uncertainties may be required to better interpret the reported constraints.
\section*{Acknowledgements}

G.T. aknowleges the financial support from  FAPERJ (PhDmerit fellowship - FAPERJ NOTA 10, 02.432/2024).
The authors aknowledges Dr. Griffin Hosseinzadeh for valuable discussions and insights on shock cooling modeling.

C.R.B. acknowledges the financial support from CNPq (316072/2021-4) and from FAPERJ (grants 201.456/2022 and 210.330/2022) and the FINEP contract 01.22.0505.00 (ref. 1891/22). 
The authors made use of Sci-Mind servers machines developed by the CBPF AI LAB team and would like to thank P. Russano and M. Portes de Albuquerque for all the support in infrastructure matters.
C.D.K. gratefully acknowledges support from the NSF through AST-2432037, the HST Guest Observer Program through HST-SNAP-17070 and HST-GO-17706, and from JWST Archival Research through JWST-AR-6241 and JWST-AR-5441.

P.K.H. gratefully acknowledges the Fundação de Amparo à Pesquisa do Estado de São Paulo (FAPESP) for the support grant 2023/14272-4.

This work includes data obtained with the Swope Telescope at Las Campanas Observatory, Chile, as part of the Swope Time Domain Key Project (PI: Piro, Co-Is: Drout, Phillips, Holoien, Burns, Madore, Foley, Coulter, Rojas-Bravo, Dimitriadis, Kilpatrick, Hsiao). We thank Jorge Anais, Abdo Campillay, and Yilin Kong Riveros for their valuable Swope observations.
Parts of this research were supported by the Australian Research Council Centre of Excellence for Gravitational Wave Discovery (OzGrav), through project number CE230100016.
AAC acknowledges financial support from the Severo Ochoa grant CEX2021-001131-S funded by MCIN/AEI/10.13039/501100011033 and the project PID2023-153123NB-I00 funded by MCIN/AEI.

Some of the data presented herein were obtained at Keck Observatory, which is a private 501(c)3 non-profit organization operated as a scientific partnership among the California Institute of Technology, the University of California, and the National Aeronautics and Space Administration. The Observatory was made possible by the generous financial support of the W.\ M.\ Keck Foundation.

The authors wish to recognize and acknowledge the very significant cultural role and reverence that the summit of Maunakea has always had within the Native Hawaiian community. We are most fortunate to have the opportunity to conduct observations from this mountain.

\appendix
\section{Photometric observations of SN 2022acko}
\label{sec:appendix_photometry}

Here we provide the novel photometric data used in this work. Table \ref{tab:appendix_photometry_table_2022acko} contains magnitude values of new observations of SN 2022acko. All of the reported magnitudes are in the AB magnitude system.  We note that all

\begin{center}
\begin{longtable}{cccc}
\caption{Photometric observations from the 2022acko dataset.}
\label{tab:appendix_photometry_table_2022acko}\\
\hline
\toprule
\textbf{MJD} & \textbf{Band} & \textbf{$m$} & \textbf{$\sigma_{m}$} \\
\midrule
\endfirsthead

\toprule
\textbf{MJD} & \textbf{Band} & \textbf{$m$} & \textbf{$\sigma_{m}$} \\
\midrule
\endhead

\midrule
\multicolumn{4}{r}{{Continue...}} \\
\endfoot

\hline
\endlastfoot

60101.185 & LCO/Sinistro $r$ & 18.440 & 0.074 \\
60101.187 & LCO/Sinistro $i$ & 18.543 & 0.115 \\
60134.745 & LCO/Sinistro $i$ & 18.863 & 0.055 \\
60134.784 & LCO/Sinistro $r$ & 18.748 & 0.027 \\
60134.788 & LCO/Sinistro $i$ & 18.909 & 0.069 \\
60139.387 & LCO/Sinistro $g$ & 19.807 & 0.072 \\
60139.388 & LCO/Sinistro $r$ & 18.881 & 0.044 \\
60139.389 & LCO/Sinistro $i$ & 19.018 & 0.058 \\
60150.054 & LCO/Sinistro $r$ & 18.929 & 0.022 \\
60150.058 & LCO/Sinistro $i$ & 19.051 & 0.037 \\
60152.809 & LCO/Sinistro $r$ & 18.933 & 0.037 \\
60152.812 & LCO/Sinistro $i$ & 19.137 & 0.049 \\
60155.410 & LCO/Sinistro $g$ & 19.997 & 0.080 \\
60155.411 & LCO/Sinistro $r$ & 19.009 & 0.045 \\
60155.412 & LCO/Sinistro $i$ & 19.157 & 0.069 \\
60199.077 & LCO/Sinistro $r$ & 19.413 & 0.034 \\
60199.080 & LCO/Sinistro $i$ & 19.640 & 0.044 \\
60204.550 & LCO/Sinistro $i$ & 19.725 & 0.071 \\
60204.562 & LCO/Sinistro $g$ & 20.325 & 0.070 \\
60204.565 & LCO/Sinistro $r$ & 19.480 & 0.038 \\
60204.568 & LCO/Sinistro $i$ & 19.788 & 0.051 \\
60224.111 & LCO/Sinistro $r$ & 19.582 & 0.080 \\
60229.095 & LCO/Sinistro $r$ & 19.668 & 0.045 \\
60229.097 & LCO/Sinistro $i$ & 19.991 & 0.074 \\
60233.715 & LCO/Sinistro $r$ & 19.737 & 0.070 \\
60233.825 & LCO/Sinistro $r$ & 19.781 & 0.077 \\
60233.826 & LCO/Sinistro $i$ & 20.061 & 0.080 \\
60252.331 & LCO/Sinistro $r$ & 19.962 & 0.127 \\
60266.314 & LCO/Sinistro $g$ & 20.753 & 0.100 \\
60266.317 & LCO/Sinistro $i$ & 20.263 & 0.127 \\
60111.422 & Swope $r$ & 18.595 & 0.013 \\
60111.429 & Swope $i$ & 18.690 & 0.026 \\
60111.437 & Swope $g$ & 19.607 & 0.044 \\
60118.394 & Swope $r$ & 18.731 & 0.045 \\
60118.402 & Swope $i$ & 18.789 & 0.036 \\
60118.410 & Swope $g$ & 19.691 & 0.035 \\
60118.418 & Swope $r$ & 18.671 & 0.011 \\
60118.426 & Swope $i$ & 18.830 & 0.017 \\
60118.433 & Swope $g$ & 19.733 & 0.036 \\
60119.381 & Swope $r$ & 18.655 & 0.010 \\
60119.389 & Swope $i$ & 18.829 & 0.016 \\
60119.397 & Swope $g$ & 19.718 & 0.025 \\
60119.405 & Swope $V$ & 19.543 & 0.020 \\
60119.413 & Swope $B$ & 20.768 & 0.049 \\
60133.354 & Swope $r$ & 18.830 & 0.023 \\
60133.362 & Swope $i$ & 18.955 & 0.025 \\
60133.369 & Swope $g$ & 19.768 & 0.058 \\
60133.377 & Swope $V$ & 19.678 & 0.054 \\
60133.384 & Swope $B$ & 21.168 & 1.086 \\
60133.403 & Swope $r$ & 18.826 & 0.017 \\
60133.410 & Swope $i$ & 19.013 & 0.024 \\
60133.418 & Swope $g$ & 19.793 & 0.044 \\
60133.426 & Swope $V$ & 19.700 & 0.039 \\
60133.434 & Swope $B$ & 21.276 & 1.086 \\
60141.383 & Swope $r$ & 18.903 & 0.013 \\
60141.391 & Swope $i$ & 19.017 & 0.017 \\
60141.398 & Swope $g$ & 19.886 & 0.022 \\
60141.406 & Swope $V$ & 19.685 & 0.022 \\
60141.414 & Swope $B$ & 20.993 & 0.041 \\
60141.425 & Swope $u$ & 22.061 & 1.086 \\
60149.407 & Swope $r$ & 18.952 & 0.016 \\
60149.422 & Swope $i$ & 19.127 & 0.020 \\
60149.430 & Swope $g$ & 19.990 & 0.041 \\
60149.438 & Swope $u$ & 21.021 & 1.086 \\
60151.397 & Swope $r$ & 18.905 & 0.013 \\
60151.405 & Swope $i$ & 19.125 & 0.025 \\
60151.413 & Swope $g$ & 19.908 & 0.032 \\
60151.420 & Swope $u$ & 21.533 & 1.086 \\
60151.433 & Swope $V$ & 19.888 & 0.042 \\
60151.441 & Swope $B$ & 20.287 & 1.086 \\
60174.308 & Swope $r$ & 19.138 & 0.016 \\
60174.315 & Swope $i$ & 19.252 & 0.021 \\
60174.323 & Swope $g$ & 20.060 & 0.031 \\
60174.332 & Swope $V$ & 19.893 & 0.032 \\
60174.345 & Swope $B$ & 20.968 & 0.053 \\
60179.338 & Swope $r$ & 19.173 & 0.016 \\
60179.349 & Swope $i$ & 19.267 & 0.027 \\
60179.358 & Swope $g$ & 20.041 & 0.036 \\
60179.366 & Swope $V$ & 19.977 & 0.036 \\
60179.373 & Swope $B$ & 21.737 & 1.086 \\
60181.284 & Swope $r$ & 19.193 & 0.017 \\
60181.291 & Swope $i$ & 19.402 & 0.025 \\
60181.299 & Swope $g$ & 20.068 & 0.031 \\
60181.306 & Swope $V$ & 19.920 & 0.034 \\
60181.314 & Swope $B$ & 20.920 & 0.058 \\
60182.327 & Swope $r$ & 19.243 & 0.013 \\
60182.342 & Swope $i$ & 19.381 & 0.024 \\
60182.349 & Swope $g$ & 20.106 & 0.025 \\
60182.357 & Swope $V$ & 20.018 & 0.035 \\
60182.365 & Swope $B$ & 20.896 & 0.056 \\
60195.354 & Swope $r$ & 19.354 & 0.013 \\
60195.365 & Swope $i$ & 19.547 & 0.024 \\
60195.376 & Swope $g$ & 20.243 & 0.047 \\
59933.047 & CTIO/S-PLUS $g$ & 16.250 & 0.013 \\
59939.129 & CTIO/S-PLUS $g$ & 16.361 & 0.007 \\
59939.131 & CTIO/S-PLUS $r$ & 16.072 & 0.006 \\
59939.132 & CTIO/S-PLUS $i$ & 16.200 & 0.008 \\
59939.133 & CTIO/S-PLUS $z$ & 16.228 & 0.013 \\ 
59943.110 & CTIO/S-PLUS $r$ & 16.104 & 0.006 \\ 
59943.111 & CTIO/S-PLUS $i$ & 16.180 & 0.008 \\
59943.112 & CTIO/S-PLUS $z$ & 16.191 & 0.011 \\
60229.218 & CTIO/S-PLUS $g$ & 20.684 & 0.257 \\
60229.220 & CTIO/S-PLUS $g$ & 20.688 & 0.250 \\
60229.222 & CTIO/S-PLUS $r$ & 19.799 & 0.180 \\
60229.224 & CTIO/S-PLUS $r$ & 19.819 & 0.198 \\
60229.226 & CTIO/S-PLUS $i$ & 19.853 & 0.160 \\
60229.228 & CTIO/S-PLUS $i$ & 19.879 & 0.204 \\
60229.230 & CTIO/S-PLUS $z$ & 19.545 & 0.138 \\
60229.232 & CTIO/S-PLUS $z$ & 19.492 & 0.117 \\
60244.109 & CTIO/S-PLUS $r$ & 19.754 & 0.140 \\
60244.111 & CTIO/S-PLUS $r$ & 19.754 & 0.157 \\
60244.113 & CTIO/S-PLUS $i$ & 19.817 & 0.148 \\
60244.115 & CTIO/S-PLUS $i$ & 19.902 & 0.183 \\
60244.119 & CTIO/S-PLUS $z$ & 19.712 & 0.204 \\
60247.226 & CTIO/S-PLUS $r$ & 20.063 & 0.227 \\
60247.228 & CTIO/S-PLUS $r$ & 19.911 & 0.225 \\
60247.230 & CTIO/S-PLUS $i$ & 19.975 & 0.192 \\
60247.231 & CTIO/S-PLUS $i$ & 19.928 & 0.236 \\
60247.233 & CTIO/S-PLUS $z$ & 19.346 & 0.230 \\
60247.235 & CTIO/S-PLUS $z$ & 19.773 & 0.236 \\
60251.252 & CTIO/S-PLUS $g$ & 20.788 & 0.154 \\
60251.256 & CTIO/S-PLUS $g$ & 20.738 & 0.153 \\
60251.260 & CTIO/S-PLUS $r$ & 19.897 & 0.075 \\
60251.264 & CTIO/S-PLUS $r$ & 19.914 & 0.080 \\
60251.268 & CTIO/S-PLUS $i$ & 20.177 & 0.097 \\
60251.272 & CTIO/S-PLUS $i$ & 20.058 & 0.092 \\
60251.276 & CTIO/S-PLUS $z$ & 19.613 & 0.090 \\
60251.280 & CTIO/S-PLUS $z$ & 19.746 & 0.102 \\
60253.307 & CTIO/S-PLUS $g$ & 20.828 & 0.283 \\
60253.311 & CTIO/S-PLUS $g$ & 20.911 & 0.275 \\
60253.315 & CTIO/S-PLUS $r$ & 20.050 & 0.177 \\
60253.319 & CTIO/S-PLUS $r$ & 20.097 & 0.161 \\
60253.323 & CTIO/S-PLUS $i$ & 20.173 & 0.113 \\
60253.328 & CTIO/S-PLUS $i$ & 20.083 & 0.116 \\
60253.332 & CTIO/S-PLUS $z$ & 20.057 & 0.148 \\
60253.336 & CTIO/S-PLUS $z$ & 19.934 & 0.128 \\
60255.239 & CTIO/S-PLUS $g$ & 20.876 & 0.243 \\
60255.243 & CTIO/S-PLUS $g$ & 20.962 & 0.225 \\
60255.247 & CTIO/S-PLUS $r$ & 20.036 & 0.110 \\
60255.252 & CTIO/S-PLUS $r$ & 20.004 & 0.084 \\
60255.255 & CTIO/S-PLUS $i$ & 20.093 & 0.118 \\
60255.260 & CTIO/S-PLUS $i$ & 20.111 & 0.116 \\
60255.264 & CTIO/S-PLUS $z$ & 19.890 & 0.130 \\
60255.268 & CTIO/S-PLUS $z$ & 19.717 & 0.106 \\
60259.258 & CTIO/S-PLUS $g$ & 20.931 & 0.126 \\
60259.262 & CTIO/S-PLUS $g$ & 20.881 & 0.107 \\
60259.266 & CTIO/S-PLUS $r$ & 19.940 & 0.066 \\
60259.270 & CTIO/S-PLUS $r$ & 19.963 & 0.069 \\
60259.274 & CTIO/S-PLUS $i$ & 20.166 & 0.093 \\
60259.278 & CTIO/S-PLUS $i$ & 20.149 & 0.090 \\
60259.282 & CTIO/S-PLUS $z$ & 19.714 & 0.095 \\
60259.286 & CTIO/S-PLUS $z$ & 19.997 & 0.133 \\
60261.100 & CTIO/S-PLUS $g$ & 20.781 & 0.221 \\
60261.104 & CTIO/S-PLUS $g$ & 20.919 & 0.253 \\
60261.108 & CTIO/S-PLUS $r$ & 20.237 & 0.141 \\
60261.112 & CTIO/S-PLUS $r$ & 20.164 & 0.154 \\
60261.116 & CTIO/S-PLUS $i$ & 20.188 & 0.192 \\
60261.120 & CTIO/S-PLUS $i$ & 20.143 & 0.156 \\
60261.124 & CTIO/S-PLUS $z$ & 20.088 & 0.190 \\
60261.128 & CTIO/S-PLUS $z$ & 20.076 & 0.163 \\
\end{longtable}
\end{center}

\section{Impact of missing data in SN 2021yja}

\begin{figure}[h]
    \centering
    \includegraphics[width=1\linewidth]{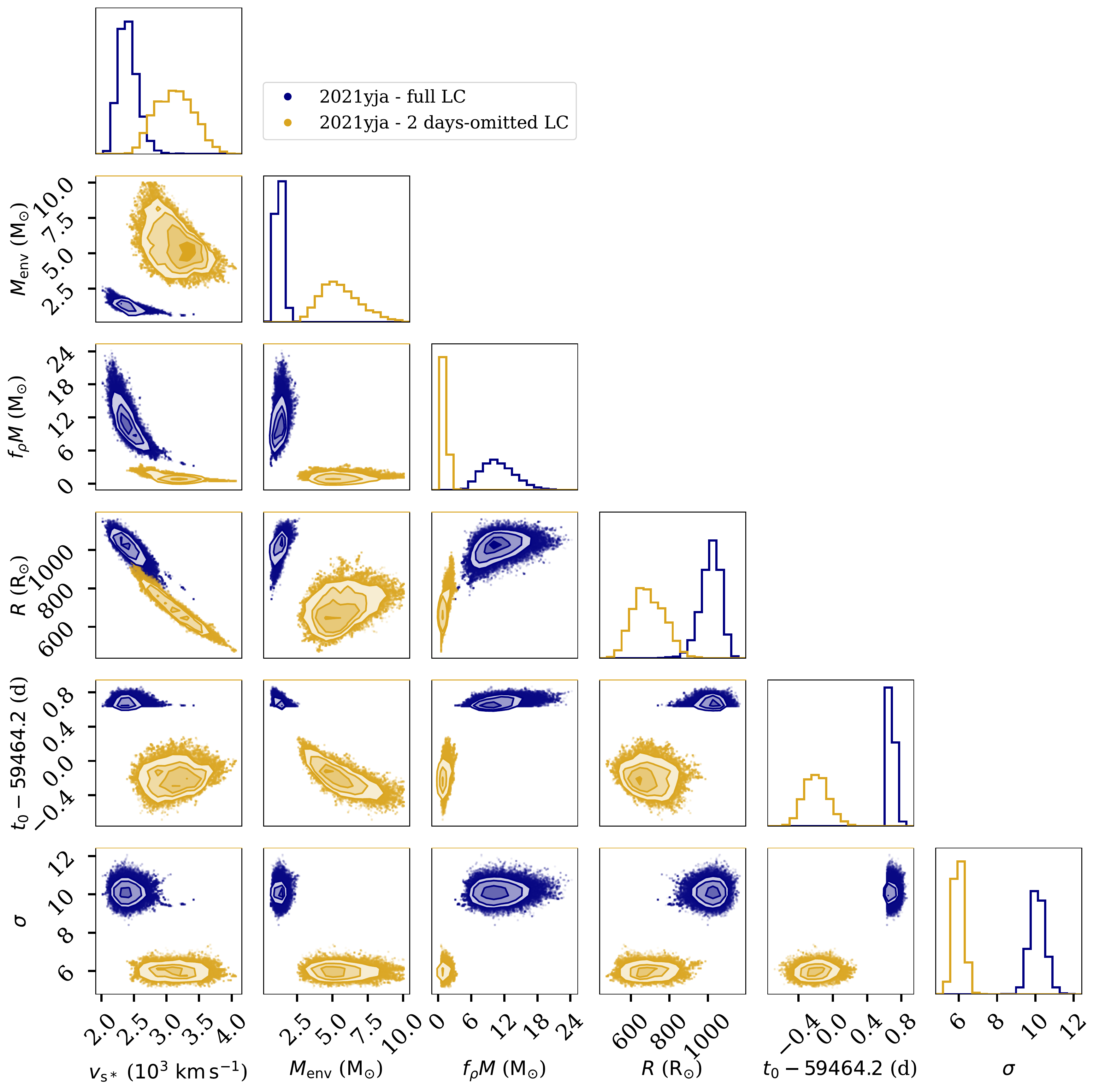}
    \caption{Posterior estimation for the MSW23 model over the early light curve of SN~2021yja.  
The blue distributions represent the results using the full multi-band light curve from \citet{Hosseinzadeh2022_2021yja}. ZThe golden distributions correspond to the estimation where the first day of observation was omitted from the light curve.}
    \label{fig:2021yjacorner}
\end{figure}

Here we present Figure~\ref{fig:2021yjacorner} as a test to estimate the potential impact of missing early-time data on shock-cooling modeling, using SN~2021yja as a case study.
It provides a complement to the discussion in Section~\ref{sec:discussion}, although it does not has any effect on the conclusions drawn in the main text.  

\bibliographystyle{aasjournal} 
\bibliography{references}

\end{document}